\documentclass[12pt]{spieman}  
\usepackage{amsmath,amsfonts,amssymb}
\usepackage{graphicx}
\usepackage{setspace}
\usepackage{tocloft}
\usepackage{siunitx}
\usepackage{booktabs}
\usepackage{dcolumn}

\title{Development of a near-infrared wide-field integral field unit by ultra-precision diamond cutting}

\author[a*]{Kosuke Kushibiki}
\author[b,c]{Shinobu Ozaki}
\author[d]{Masahiro Takeda}
\author[d]{Takuya Hosobata}
\author[d]{Yutaka Yamagata}
\author[e]{Shinya Morita}
\author[b]{Toshihiro Tsuzuki}
\author[f,g]{Keiichi Nakagawa}
\author[g]{Takao Saiki}
\author[g]{Yutaka Ohtake}
\author[b]{Kenji Mitsui}
\author[h]{Hirofumi Okita}
\author[a]{Yutaro Kitagawa}
\author[a]{Yukihiro Kono}
\author[b,c,i]{Kentaro Motohara}
\author[j]{Hidenori Takahashi}
\author[a]{Masahiro Konishi}
\author[a]{Natsuko Kato}
\author[a]{Shuhei Koyama}
\author[b,i]{Nuo Chen}

\affil[a]{Institute of Astronomy, Graduate School of Science, The University of Tokyo, 2-21-1 Osawa, Mitaka, Tokyo 181-0015, Japan}
\affil[b]{Advanced Technology Center, National Astronomical Observatory of Japan, 2-21-1 Osawa, Mitaka, Tokyo 181-8588, Japan}
\affil[c]{Department of Astronomical Science, SOKENDAI, 2-21-1 Osawa, Mitaka, Tokyo 181-8588, Japan}
\affil[d]{RIKEN Center for Advanced Photonics (RAP), RIKEN, 2-1 Hirosawa, Wako, Saitama 351-0198, Japan}
\affil[e]{Department of Advanced Machinery Engineering, School of Engineering, Tokyo Denki University, 5 Senju Asahi-cho, Adachi-ku, Tokyo 120-8551, Japan}
\affil[f]{Department of Bioengineering, Graduate School of Engineering, The University of Tokyo, 7-3-1 Hongo, Bunkyo-ku, Tokyo 113-0033, Japan}
\affil[g]{Department of Precision Engineering, Graduate School of Engineering, The University of Tokyo, 7-3-1 Hongo, Bunkyo-ku, Tokyo 113-0033, Japan}
\affil[h]{Subaru Telescope, National Astronomical Observatory of Japan, 650 North A'ohoku Place, Hilo, HI 96720, USA}
\affil[i]{Department of Astronomy, Graduate School of Science, The University of Tokyo, 7-3-1 Hongo, Bunkyo-ku, Tokyo 113-0033, Japan}
\affil[j]{Kiso Observatory, Institute of Astronomy, Graduate School of Science, The University of Tokyo, 10762-30 Mitake, Kiso-machi, Kiso-gun, Nagano 397-0101, Japan}

\cftpagenumbersoff{figure}
\cftpagenumbersoff{table} 
\begin{document} 
\maketitle

\begin{abstract}
Integral Field Spectroscopy (IFS) is an observational method to obtain spatially resolved spectra over a specific field of view (FoV) in a single exposure. In recent years, near-infrared IFS has gained importance in observing objects with strong dust attenuation or at high redshift. One limitation of existing near-infrared IFS instruments is their relatively small FoV, less than 100~\si{arcsec^2}, compared to optical instruments. Therefore, we have developed a near-infrared (0.9--2.5~\si{\um}) image-slicer type integral field unit (IFU) with a larger FoV of 13.5 $\times$ 10.4~\si{arcsec^2} by matching a slice width to a typical seeing size of 0.4~\si{arcsec}. The IFU has a compact optical design utilizing off-axis ellipsoidal mirrors to reduce aberrations. Complex optical elements were fabricated using an ultra-precision cutting machine to achieve RMS surface roughness of less than 10~\si{nm} and a P-V shape error of less than 300~\si{nm}. The ultra-precision machining can also simplify alignment procedures. The on-sky performance evaluation confirmed that the image quality and the throughput of the IFU were as designed. In conclusion, we have successfully developed a compact IFU utilizing an ultra-precision cutting technique, almost fulfilling the requirements.
\end{abstract}


{\noindent \footnotesize\textbf{*}Kosuke Kushibiki,  \linkable{k.kushibiki@ioa.s.u-tokyo.ac.jp} }

\section{Introduction}
Integral Field Spectroscopy (IFS) is an observational method to obtain spatially resolved spectra over a specific field of view (FoV) in a single exposure. It can obtain spectral information of spatially extended objects more efficiently than conventional scanning methods, such as slit scanning, scanning Fabry-Perot interferometry, and imaging Fourier transform spectroscopy. It also allows us to obtain homogeneous data over the FoV because it is not affected by temporal variations in atmospheric and instrument conditions. Therefore, many IFS instruments have been developed for the optical and infrared astronomy \cite{Bacon2010, Morrissey2018, Allington-Smith2002, Ozaki2020, Eisenhauer2003, Sharples2013, Larkin2006, Allington-Smith2006a, McGregor2003, Closs2008}.

In particular, near-infrared IFS is important to study nearby objects with strong dust attenuation or to obtain rest-frame UV and optical spectra of high-redshift objects. However, as the FoVs of existing near-infrared IFS instruments for 8--10~\si{m} telescopes are smaller than 100~\si{arcsec^2}\cite{Eisenhauer2003, Sharples2013, Larkin2006, Allington-Smith2006a, McGregor2003, Closs2008}, various science cases that require wider FoV, such as observation of extended objects or wide-field surveys, are restricted. This is because they are optimized for observation with high spatial resolution using adaptive optics (AO).

Therefore, we have developed a near-infrared image-slicer type integral field unit (IFU) with a large FoV for an existing near-infrared instrument, Simultaneous-color Wide-field Infrared Multi-object Spectrograph (SWIMS)\cite{Konishi2010, Konishi2012, Motohara2014, Motohara2016, Konishi2018, Konishi2020}. SWIMS is one of the first generation instruments for the University of Tokyo Atacama Observatory (TAO) 6.5~\si{m} telescope, which is being constructed by the Institute of Astronomy, the University of Tokyo, on the summit of Cerro Chajanantor in the Atacama Desert, Chile\cite{Yoshii2010, Yoshii2014, Yoshii2016, Doi2018, Yoshii2020, Miyata2022}. SWIMS is capable of simultaneous two-color imaging with a large FoV of $\phi$~9.6~\si{arcmin} and multi-object slit spectroscopy covering the entire 0.9--2.5~\si{\um} in a single exposure. 

In this paper, we report the development and performance evaluation of the IFU for SWIMS, which is named SWIMS-IFU. A basic concept of the IFU is explained in Section~\ref{sec: concept} and its optical design is presented in Section~\ref{sec: optics}. Section~\ref{sec: manufacturing} summarizes mirror fabrication and assembly using ultra-precision cutting. After showing performance evaluation by laboratory tests in Section~\ref{sec: lab_evaluation} and an engineering observation in Section~\ref {sec: eng_observation}, we summarize the overall development in Section~\ref {sec: conclusions}. 

\section{Concept}
\label{sec: concept}
\subsection{Overview of SWIMS}
\begin{figure}[tb]
    \centering
    \includegraphics[keepaspectratio, width=0.6\hsize]{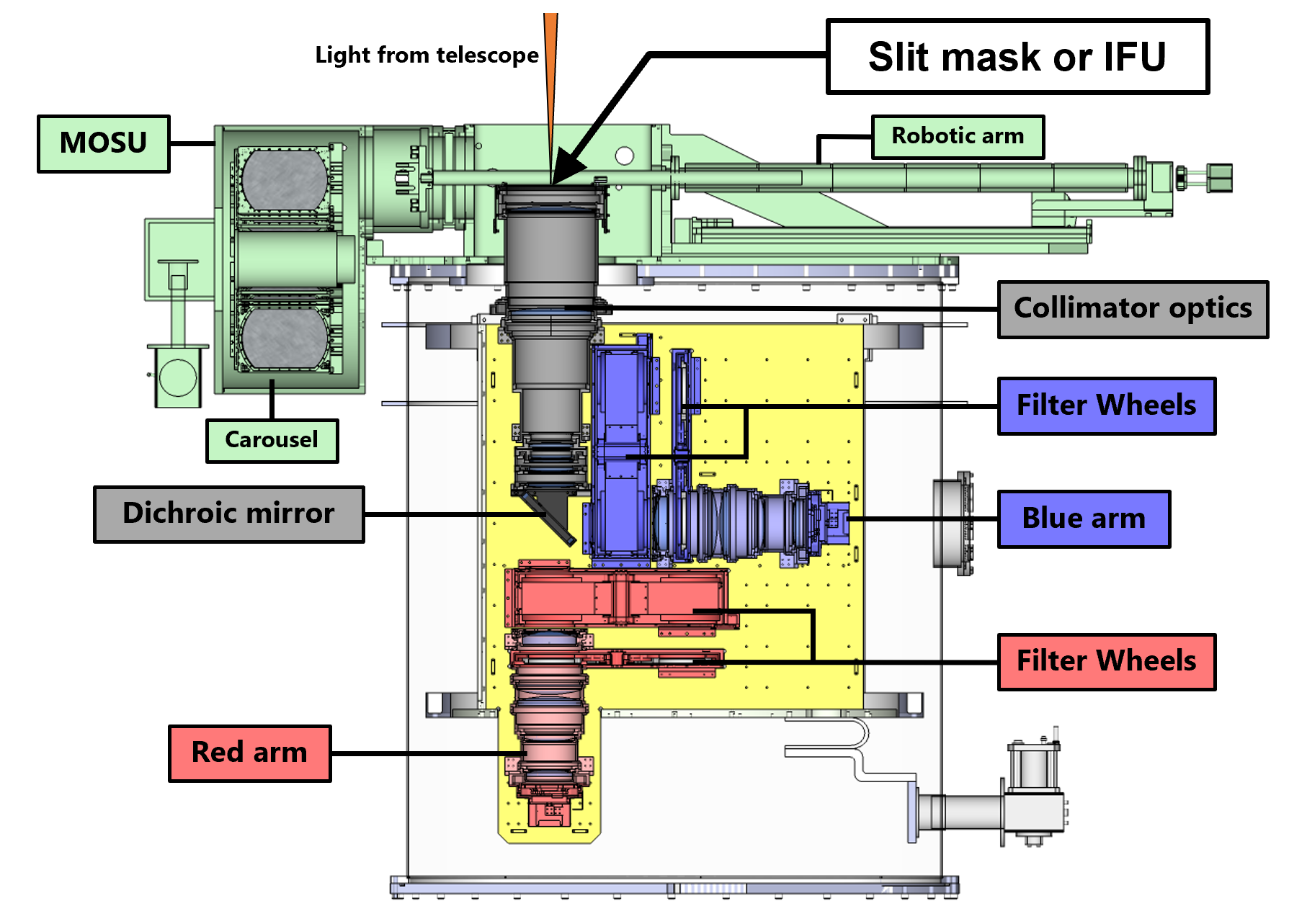}
    \caption{Internal structure of SWIMS. A slit mask or the IFU is placed on the telescope focal plane indicated by the arrow.}
    \label{fig:swims_mechanics}
\end{figure}

Internal structure of SWIMS is shown in Figure~\ref{fig:swims_mechanics}. Imaging with a large FoV and spectroscopy with a wide simultaneous wavelength coverage of 0.9--2.5~\si{\um} are realized by dividing collimated light into two wavelength ranges of 0.9--1.45~\si{\um} (blue arm) and 1.45--2.5~\si{\um} (red arm) by a dichroic mirror and by re-imaging them simultaneously on focal plane array detectors of each arm. Switching between imaging and spectroscopy modes is done by changing filters and grisms in filter wheels. Two 2048 $\times$ 2048 format HAWAII-2RG (H2RG) array detectors for each focal plane cover 6.7 $\times$ 3.3~\si{arcsec^2} FoV, and the goal is to increase them to four in a 2 $\times$ 2 matrix to cover the whole FoV delivered by the optics. Spectra are dispersed over the two detectors to cover the wide wavelength range.

For multi-object spectroscopy, a slit mask is placed at a telescope focal plane by a Multi-Object Spectroscopy Unit (MOSU)\cite{Takahashi2014}. The MOSU consists of a carousel that stores up to 22 slit masks under vacuum and cryogenic conditions and a robotic arm that handles the slit masks.

SWIMS will be installed at the Nasmyth focus of the TAO 6.5~{\si{m}} telescope which has a focal ratio (F-ratio) of 12.2. This is the same as the F-ratio of the Cassegrain focus of the Subaru telescope. SWIMS was carried to the Subaru Telescope to evaluate and optimize its on-sky performance before the completion of the TAO 6.5~{\si{m}} telescope. Engineering observations were conducted between 2018 and 2020. Thereafter, the instrument was operated as a PI-type instrument from 2021 to 2022.

\subsection{Concept of the SWIMS-IFU}
The SWIMS-IFU is an image-slicer type IFU. It should be stored and operated in a cryogenic environment to suppress thermal background radiation from the IFU itself. By placing the IFU at the telescope focal plane as slit masks, SWIMS can be switched to an IFS mode. Thus, the IFU is also stored in a dedicated slot of the carousel and installed at the telescope focal plane by the MOSU. At the telescope focal plane, the IFU is held in place by magnets embedded in both the IFU and focal plane structure, which is the same mechanism as slit masks. This basic concept of the IFU does not require any modification of the optics of SWIMS. On the other hand, the size of the IFU is restricted to within 235 $\times$ 170 $\times$ 55~\si{mm^3} due to space limitations in the carousel.

\section{Optics of the SWIMS-IFU}
\label{sec: optics}
The SWIMS-IFU aims to realize near-infrared IFS with an FoV larger than 100~\si{arcsec^2} and a wide simultaneous wavelength coverage of 0.9--2.5~\si{\um}. To realize the large FoV, we set a slice width of the IFU to be 0.4~\si{arcsec} which is the same as a typical seeing size at good observation sites. The number of slices is 26 (CH$+$13 to CH$+$1 and CH$-$1 to CH$-$13) and the length of each slice is 13.5~\si{arcsec}, resulting in the FoV of 13.5 $\times$ 10.4~\si{arcsec^2}, which is larger than that of existing instruments by more than a factor of two. Specifications of the IFU are summarized in Table~\ref{tab:ifu_spec}. Note that these specifications are for the Subaru Telescope. We also should note that only the central 12 channels (from CH$-$6 to CH$+$6) are currently covered by two H2RG detectors available in each arm of SWIMS, which corresponds to 13.5 $\times$ 4.8~{\si{arcsec^2}}. The FoV currently achieved is almost half of the full specification, but it is comparable to the FoV of 8 $\times$ 8~{\si{arcsec^2}} of VLT/ERIS, which is the largest FoV of any existing near-infrared IFS instruments. The full FoV of 13.5 $\times$ 10.4~{\si{arcsec^2}} can be achieved by installing two additional H2RG detectors on each arm.

\begin{table}[tb]
    \centering
    \caption{Specifications of the SWIMS-IFU.}
    \label{tab:ifu_spec}
    \begin{tabular}{lc} \toprule
         \rule[0ex]{0pt}{2.5ex}Field of view & 13.5 $\times$ 10.4~\si{arcsec^2} \\
         \rule[-1ex]{0pt}{2ex} & (13.5 $\times$ 4.8~{\si{arcsec^2}})$^{\ast}$\\
         \rule[-1ex]{0pt}{3.5ex}Slice width & 0.4~\si{arcsec} \\
         \rule[-1ex]{0pt}{3.5ex}Number of slices & 26 \\
         \rule[-1ex]{0pt}{3.5ex}Slice length & 13.5~\si{arcsec} \\
         \rule[-1ex]{0pt}{3.5ex}Wavelength coverage & 0.9--2.5~\si{\um} \\
         \rule[-1ex]{0pt}{3.5ex}Spectral resolution $\lambda/\Delta \lambda$ & 875--1500 (Blue arm), 750--1250 (Red arm) \\ 
         \rule[-1ex]{0pt}{3.5ex}Input \& Output F-ratio & 12.2 \\
         \rule[-1ex]{0pt}{3.5ex}Weight & 900~\si{g} \\ \bottomrule
         $^{\ast}$ FoV covered by central 12 channels. & 
    \end{tabular}
\end{table}

\begin{figure}[tb]
    \centering
    \includegraphics[keepaspectratio, width=0.6\hsize]{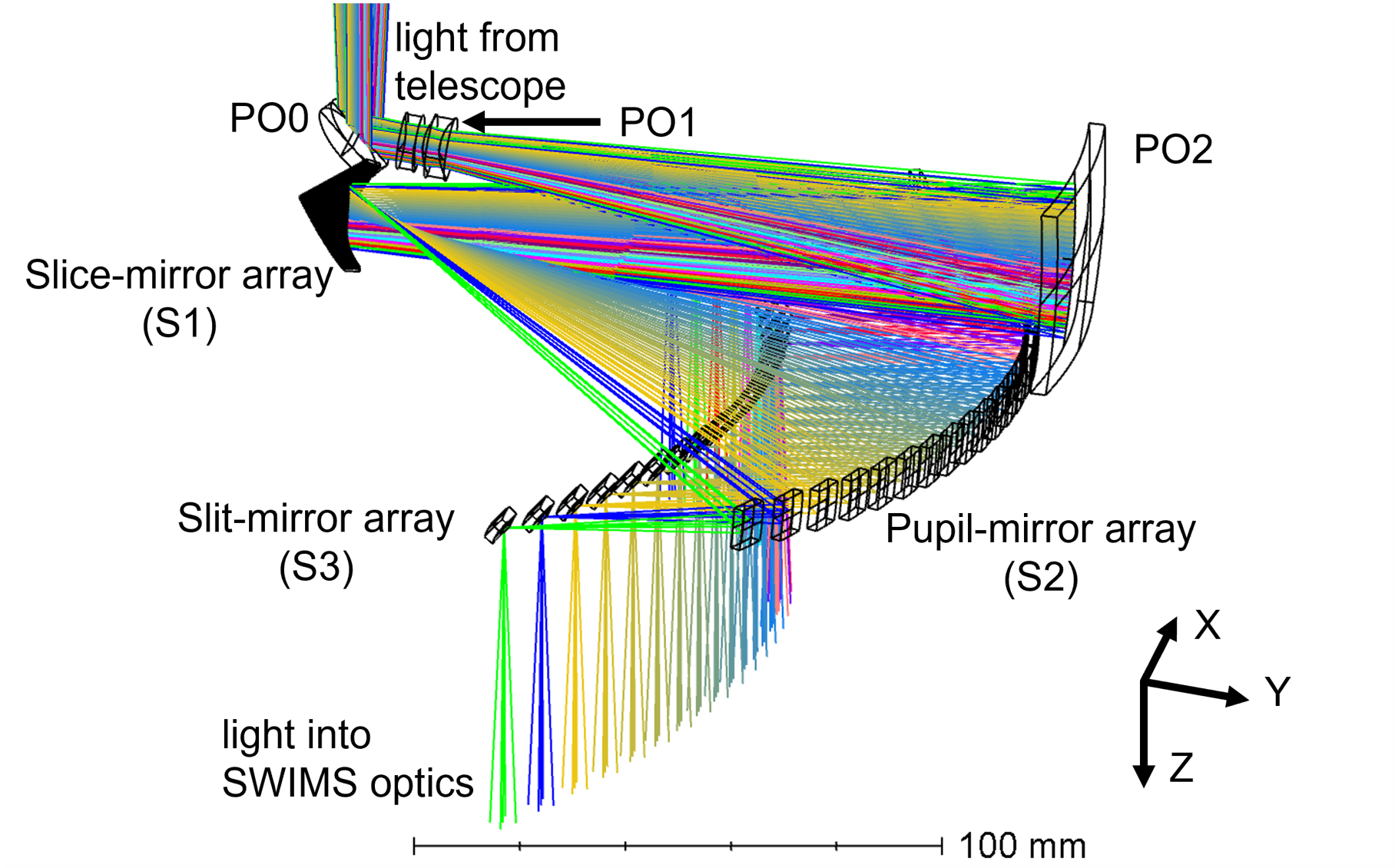}
    \caption{Optical layout of the SWIMS-IFU.}
    \label{fig:swimsifu_optics}
\end{figure}

Figure~\ref{fig:swimsifu_optics} shows an optical layout of the IFU. To realize the compact optical design, we adopt the Advanced Image Slicer configuration\cite{Content1997}, where pupil images are formed on pupil mirrors to minimize the size of pupil mirrors and to reduce the size of the whole optical system.

A prior optics consisting of PO0 (flat mirror), PO1 (diverging doublet lens), and PO2 (concave spherical mirror) magnifies an object image by a factor of 2.75 to match the slice width to the seeing size and forms it on a slice-mirror array. An exit pupil of the prior optics is located behind the slice-mirror array. The PO1 consists of a biconcave lens made of BaF$_2$ and a meniscus lens made of S-TIH14 (OHARA INC.), both with an effective diameter of 9.4~\si{mm}.

The slice-mirror array, consisting of 26 flat slice mirrors, divides the magnified object image into 26 rectangular slices. Each slice mirror has different orientations and reflects incoming light to a corresponding pupil mirror to form a pupil image on it. The 26 slice mirrors are labeled as CH$+$13 to CH$+$1 and CH$-$1 to CH$-$13 from top to bottom in Figure~\ref{fig:swimsifu_optics}. The width of the slice mirrors is 0.52~\si{mm}, corresponding to 0.4~\si{arcsec}, and the length is 18.0~\si{mm}, corresponding to 13.5~\si{arcsec}.

A pupil-mirror array forms pseudo-slit images, which are de-magnified and sliced object images, on a slit-mirror array. The magnification of the pupil mirrors is the inverse of that of the prior optics so that the overall magnification of the IFU remains unity. The pupil-mirror array consists of 12 concave spherical mirrors for central channels (CH$-$6 to CH$+$6) and 14 concave off-axis ellipsoidal mirrors for outer channels (CH$-$13 to CH$-$7 and CH$+$7 to CH$+$13). The off-axis ellipsoidal mirrors reduce aberrations caused by large reflection angles for the outer channels.

The slit-mirror array reflects light into the SWIMS optics and pupil images are formed at a cold stop of SWIMS. The slit-mirror array consists of concave spherical mirrors. The curvature and angle of each slit mirror are adjusted to form pupil images at the correct position of the cold stop.

\begin{figure}[tb]
    \centering
    \includegraphics[keepaspectratio, width=\hsize]{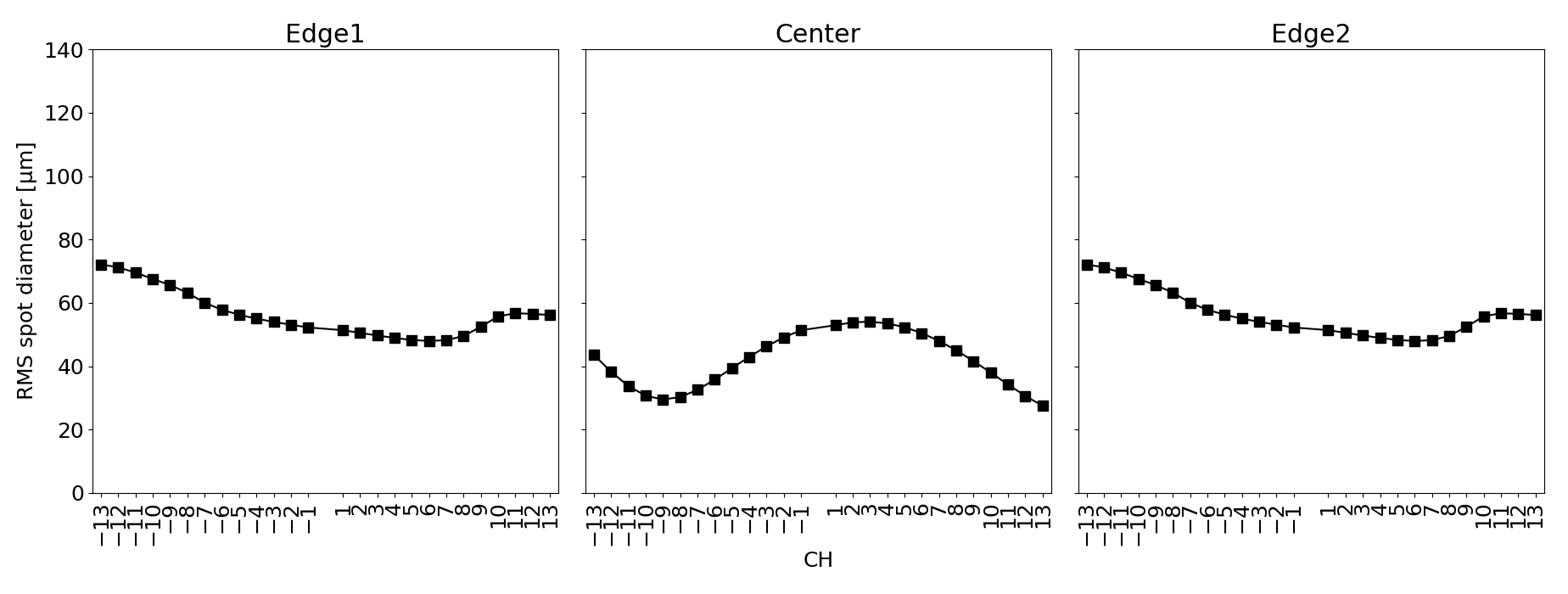}
    \caption{RMS spot diameters at the slice-mirror array for the center and both edges of each channel.}
    \label{fig:design_image_quality_slicer}
\end{figure}

\begin{figure}[tb]
    \centering
    \includegraphics[keepaspectratio, width=\hsize]{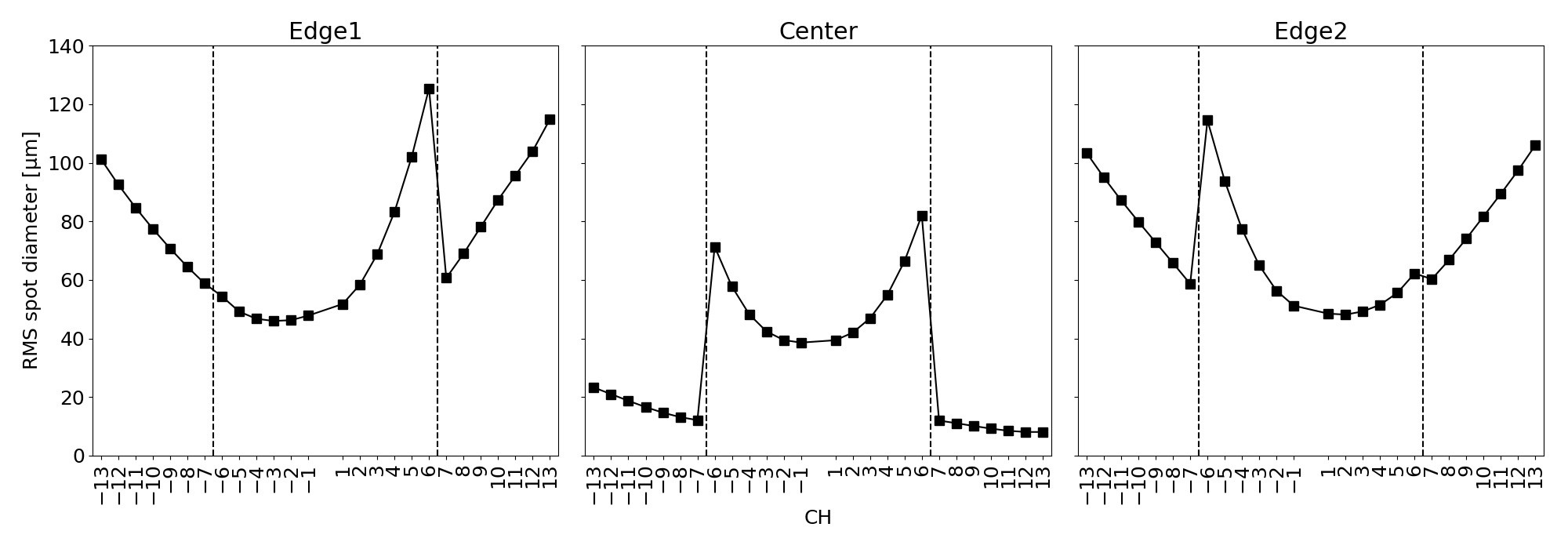}
    \caption{RMS spot diameters at the telescope focal plane at the center and both edges of each pseudo-slit. Dashed vertical lines indicate the boundaries between channels with a spherical pupil mirror and with an off-axis ellipsoidal pupil mirror. Outer channels adopt the off-axis ellipsoidal pupil mirrors.}
    \label{fig:design_image_quality}
\end{figure}

Figure~\ref{fig:design_image_quality_slicer} shows root mean square (RMS) diameters of ray-traced spots formed by the prior optics at the slice-mirror array. All the spot sizes are significantly smaller than the slice width of 0.52~\si{mm}. RMS diameters of ray-traced spots formed by the IFU at the telescope focal plane are also shown in Figure~\ref{fig:design_image_quality}.  Improvement by the off-axis ellipsoidal pupil mirrors in the outer channels is clearly seen. The largest RMS spot diameter is 125~\si{\um} at the edge of CH$+$6, which corresponds to 0.26~\si{arcsec} at the telescope focal plane.

\section{Manufacturing}
\label{sec: manufacturing}
\subsection{Mirror Fabrication by Ultra-precision Cutting}
\begin{figure}[tb]
    \centering
    \includegraphics[keepaspectratio,  width=0.6\hsize]{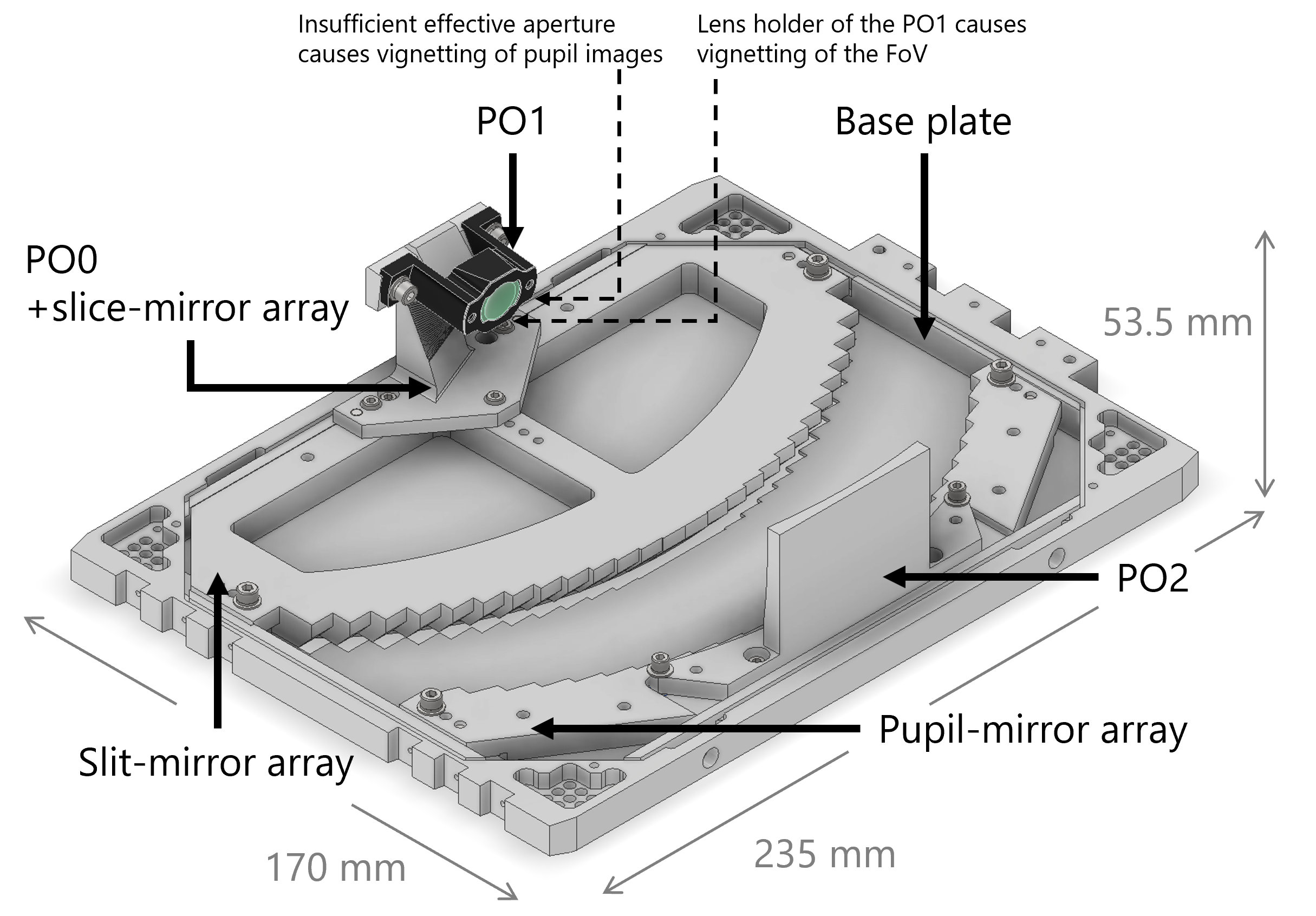}
    \caption{Mechanical structure of the SWIMS-IFU. Vignetting points of pupil images and the FoV described in Section~\ref{subsec: vignetting} are also shown.}
    \label{fig:swimsifu_mechanics}
\end{figure}

Figure~\ref{fig:swimsifu_mechanics} shows a mechanical structure of the SWIMS-IFU. It consists of five optical elements (PO0$+$slice-mirror array, PO1, PO2, pupil-mirror array, and slit-mirror array) and a base plate. All the optical surfaces except for the PO1 and interface surfaces including those of the base plate are fabricated with ultra-precision cutting. Our goal is to achieve sufficient surface quality of the mirrors only by cutting and also to guarantee relative positions and orientations between mirrors within a single optical element as well as between the elements, which will greatly save us time and effort for alignment procedures.

\begin{figure}[tp]
    \centering
    \includegraphics[keepaspectratio,  width=0.8\hsize]{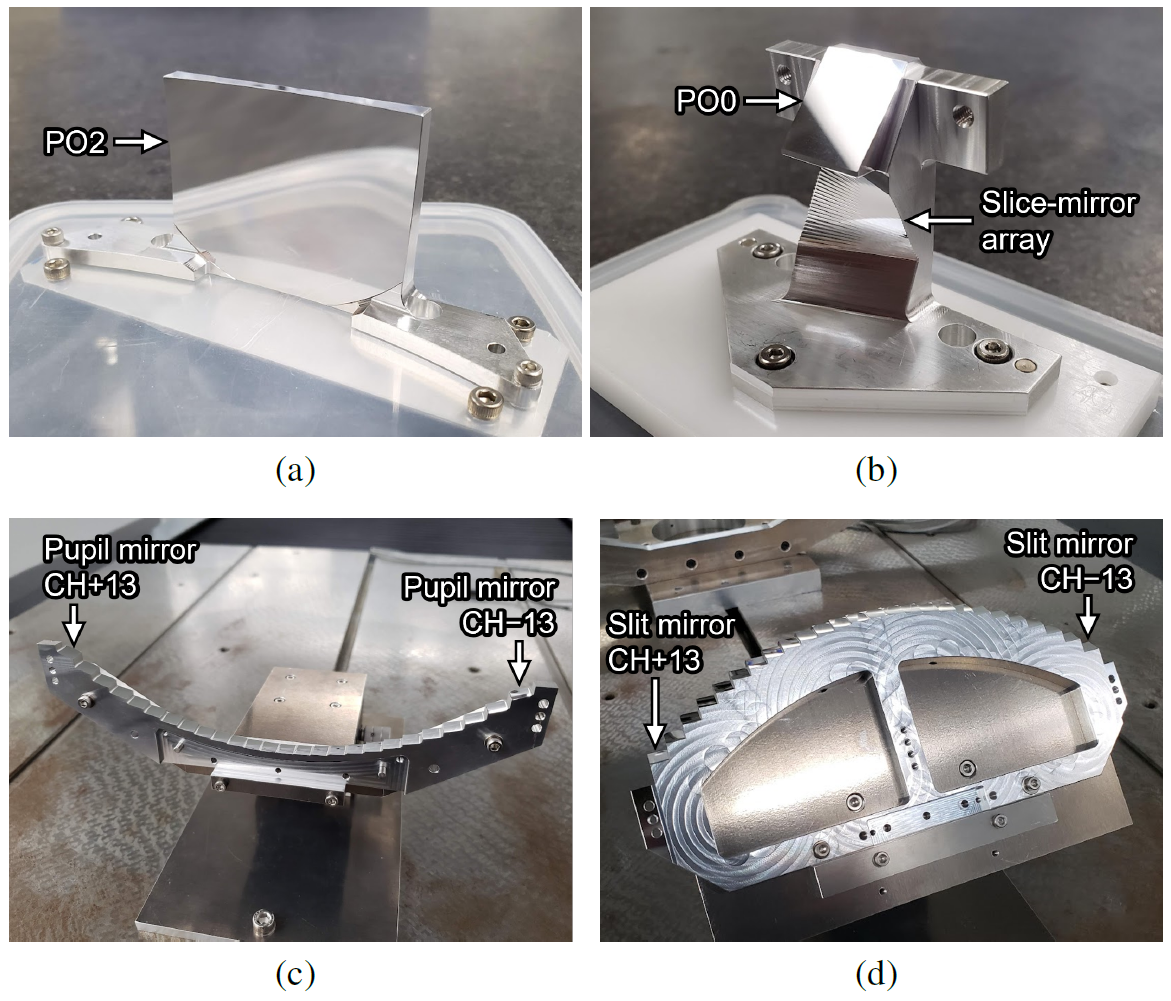}
    \caption{Completed mirrors for the SWIMS-IFU: (a) PO2, (b) PO0$+$slice-mirror array, (c) Pupil-mirror array, and (d) Slit-mirror array.}
    \label{fig:completed_mirrors}
\end{figure}

The ultra-precision cutting of the surfaces was done with ULG-100D(5A) (Shibaura Machine Co., Ltd.), operated by RIKEN, using various machining methods summarized in Table~\ref{tab:summary_of_fabrication}. All the mirrors except for the PO2 are made of a rapidly solidified aluminum alloy RSA6061\cite{Gubbels2008} to achieve a good surface roughness of less than 10~\si{nm}. The completed optical elements are shown in Figure~\ref{fig:completed_mirrors}. Unique points in our fabrication are (1) the PO0$+$slice-mirror array which has two types of flat mirrors in a single workpiece and (2) the pupil-mirror array with spherical and ellipsoidal mirrors and the slit-mirror array with various spherical mirrors. In the former fabrication, to complete two types of flat mirrors in a single operation two types of tools were mounted simultaneously on the ultra-precision machine. One of the tools is a custom-made flat diamond tool for the slice mirrors, which has the same width and cross-sectional shape as the slice mirrors\cite{Takeda2022}. The latter fabrications adopted a ball-end mill tool with a small cutting-edge radius to fabricate the spherical and ellipsoidal mirrors in a single operation. The method can process various spherical and ellipsoidal mirrors with a radius of curvature larger than that of the tool.

\begin{table}[tp]
    \centering
    \caption{Summary of fabrication by ultra-precision cutting.}
    \label{tab:summary_of_fabrication}
    \scalebox{0.75}{
    \begin{tabular}{lcccc} \toprule
         & PO2 & PO0$+$slice-mirror array & Pupil-mirror array & Slit-mirror array  \\ \midrule
        Number of mirrors & 1 & \multicolumn{1}{l}{\hspace{55pt}PO0: 1} & 26 & 26 \\
         & & \multicolumn{1}{l}{\hspace{55pt}Slice mirrors: 26} & & \vspace{0.5cm} \\
        Mirror shape & Spherical & Flat & Spherical & Spherical \\
         & & & \& Ellipsoidal & \vspace{0.5cm}\\
        Mirror size [\si{mm^2}] & $52\times41$ & \multicolumn{1}{l}{\hspace{40pt}PO0: $14\times16$} & $\sim7.25\times7$  & $\sim 7.25 \times 6.8$ \\
         & & \multicolumn{1}{l}{\hspace{40pt}Slice mirrors: $18\times0.52$} & & \vspace{0.5cm}\\
        Material & A6061 & RSA6061 & RSA6061 & RSA6061 \vspace{0.5cm}\\ 
        Fabrication method & Turning & Shaper cutting & Milling & Milling \vspace{0.5cm}\\ 
        Tool & Round-tip Tool & \multicolumn{1}{l}{\hspace{35pt}PO0: Round-tip tool} & Ball end mill & Ball end mill \\ 
         & ($R=0.5~\si{mm}$) & \multicolumn{1}{l}{\hspace{70pt}($R=2.0~\si{mm}$)} & ($R=1.0~\si{mm}$) & ($R=0.5~\si{mm}$)\\ 
         & & \multicolumn{1}{l}{\hspace{35pt}Slice mirrors: Custom-made flat tool} & &\\ 
         & & \multicolumn{1}{l}{\hspace{105pt} ($w=0.52~\si{mm}$)} & & \vspace{0.5cm}\\ 
        Finishing pitch [\si{\um}] & $\sim2.8$ & \multicolumn{1}{l}{\hspace{55pt}PO0: $\sim16.7$} & 10 & 10 \\
         & & \multicolumn{1}{l}{\hspace{55pt}Slice mirrors: Single cut} & & \\ \bottomrule
    \end{tabular}
    }
\end{table}

\subsection{Mirror Surface Quality}
\subsubsection{Surface roughness}
Surface roughness is an important indicator to assess scattering loss at a mirror surface. The scattering loss $L_{\mathrm{scatter}}$ by the five mirrors of the IFU at a wavelength $\lambda$ is estimated by the following equation\cite{Allington-Smith2006a}, 
\begin{equation}
L_{\mathrm{scatter}} = 1 - \prod_{i=1-5}\left(  1-\left(\frac{4\pi\sigma_i}{\lambda}\right)^2 \right), \label{eq:roughness}
\end{equation}
where $\sigma_i$ is RMS surface roughness of each mirror. A requirement for the total scattering loss is less than 10\% at 0.9~\si{\um}, which is translated to mean RMS surface roughness of less than 10~\si{nm} per surface when surface roughness tolerances are equally assigned to all the mirror surfaces.

\begin{figure}[tb]
    \centering
    \includegraphics[keepaspectratio,  width=0.9\hsize]{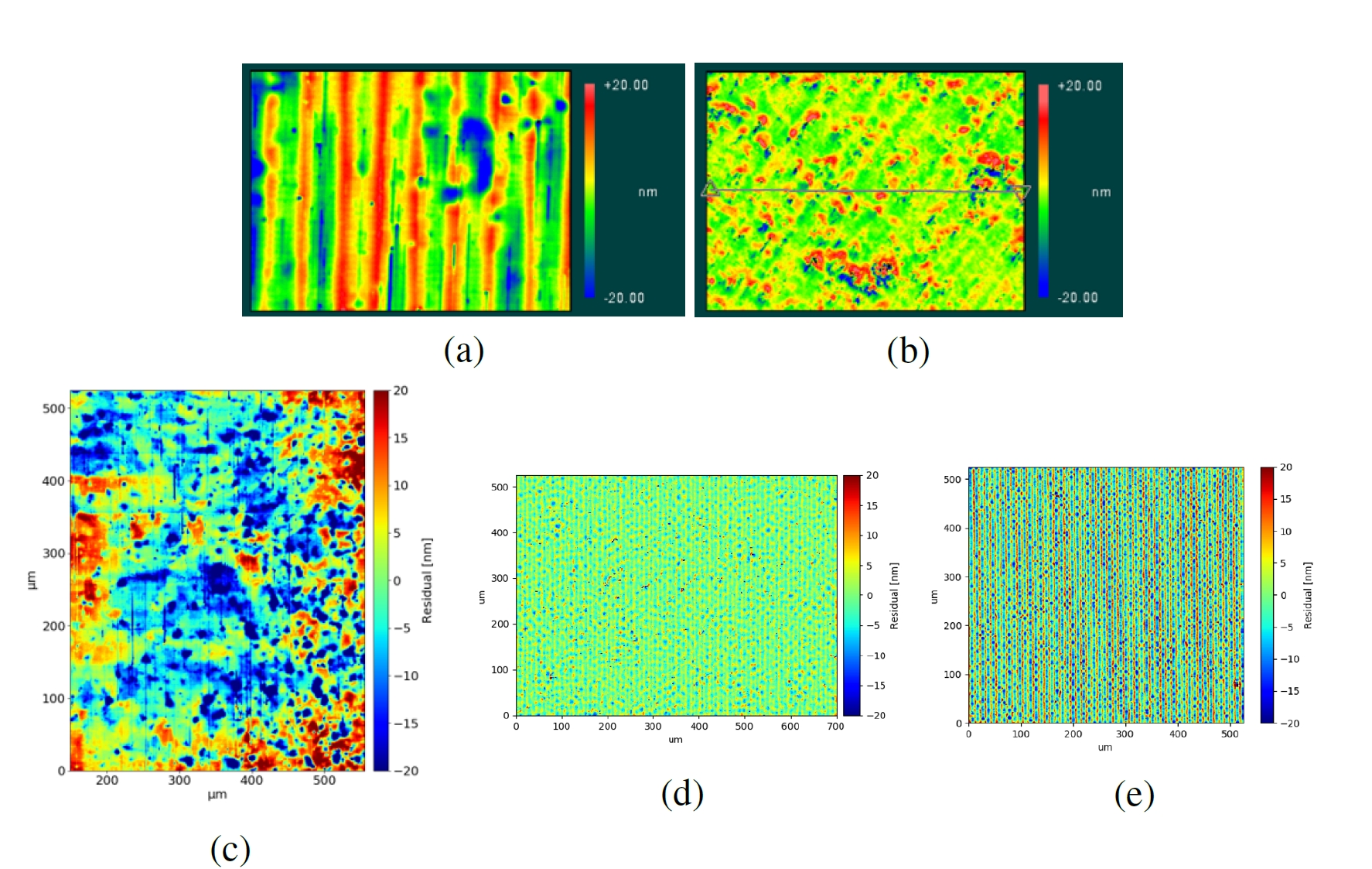}
    \caption{Measured surface images of (a) PO0, (b) PO2, (c) Slice mirror of CH$-$1, (d) Pupil mirror of CH$-$1, and (e) Slit mirror of CH$-$1.}
    \label{fig:mirror_roughness}
\end{figure}

The surface roughness of each mirror was measured using a white light interferometer (Zygo NewView 7200). Figure~\ref{fig:mirror_roughness} shows examples of measured surface images. For the PO0 and the PO2, five and six points in each mirror surface were measured, respectively, with a measurement area of 140 $\times$ 105~\si{\um^2}. For the three mirror arrays, we used a different objective lens with a long working distance not to interfere with their complex structures, which results in lower spatial sampling but a larger measurement area of 702 $\times$ 526~\si{\um^2}. Three points in each mirror were measured for the slice-mirror array and the slit-mirror array, whereas five points were measured for the pupil-mirror array. While all the mirrors were measured for the pupil-mirror array and the slit-mirror array, only 10 mirrors from CH$-$1 to CH$+$9 were measured for the slice-mirror array due to the interference between the workpiece and the objective lens. For the pupil mirrors and the slit mirrors, a low-pass filter was applied to remove patterns with a spatial period larger than 20~\si{\um} (twice the finishing pitch), which were considered to be shape errors rather than surface roughness.

As shown in Table~\ref{tab:surface_roughness}, mean RMS surface roughness, except for the slice mirrors, is 5--8~\si{nm}, which meets the requirement of less than 10~\si{nm}. The slice mirrors have the larger mean RMS surface roughness of 12.74~\si{nm}, which is due to many impurities and unsmooth patterns of the base material of RSA6061. The resulting RMS surface roughness is comparable to previous studies using a fly-cut method\cite{Allington-Smith2006a, Purll2010, Dubbeldam2012}. Combining the mean surface roughness of each mirror in Table \ref{tab:surface_roughness}, the total scattering loss of five mirrors is estimated to be 6.4\%, which still meets the requirement of scattering loss of less than 10\%.

\begin{table}[tb]
    \centering
    \caption{Measured RMS surface roughness of the mirrors.}
    \label{tab:surface_roughness}
    \begin{tabular}{lrrrrr} \toprule
        \rule[-1ex]{0pt}{3.5ex} & \multicolumn{1}{c}{PO0} & \multicolumn{1}{c}{PO2} & \multicolumn{1}{c}{Slice mirrors} & \multicolumn{1}{c}{Pupil mirrors} & \multicolumn{1}{c}{Slit mirrors} \\ \midrule
        \rule[-1ex]{0pt}{3.5ex}Mean [nm] & 8.08 & 4.90 & 12.74 & 5.50 & 7.36 \\
        \rule[-1ex]{0pt}{3.5ex}Min [nm] & 6.62 & 4.22 & 8.09 & 3.83 & 4.42 \\
        \rule[-1ex]{0pt}{3.5ex}Max [nm] & 9.43 & 5.49 & 18.84 & 9.53 & 12.80 \\ \bottomrule
    \end{tabular}
\end{table}

\subsubsection{Shape error}
Shape error is an important parameter related to image quality. In particular, the shape error of the pupil mirrors largely affects the image quality, because pupil images are formed on the pupil mirrors. Therefore, we simulated degradation of the image quality caused by the shape error of the pupil mirrors using Zemax. In this simulation, we required the degradation of an RMS spot radius at the position of the slit mirror to be less than 10\%, and we found that a peak-to-valley (P-V) shape error of each mirror should be less than 300~\si{nm}.

\begin{figure}[tb]
    \centering
    \includegraphics[keepaspectratio,  width=0.7\hsize]{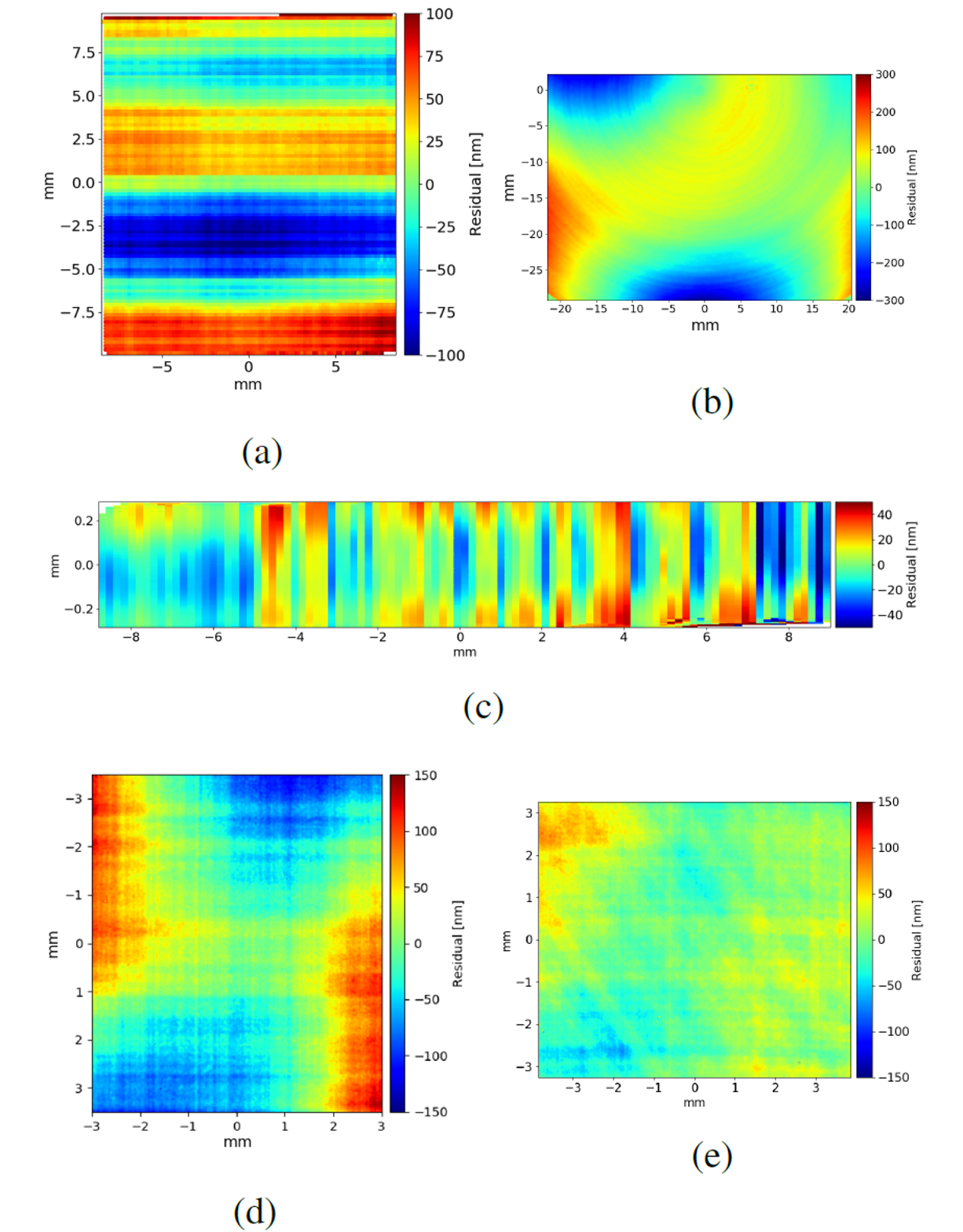}
    \caption{Measured shape error profile of (a) PO0, (b) PO2, (c) Slice mirror of CH$-$1, (d) Pupil mirror of CH$-$1, and (e) Slit mirror of CH$-$1.}
    \label{fig:mirror_shapeerror}
\end{figure}

The shape errors of the fabricated mirrors were measured with a laser interferometer (Zygo Verifire QPZ). Figure~\ref{fig:mirror_shapeerror} shows examples of the measured shape error profiles. The measurement of the slice mirrors was performed only for six channels, CH$\pm$1, $\pm$6, and $\pm$13, due to time constraints. Also, only the spherical surfaces of the pupil mirrors were measured because an aspherical surface is difficult to be measured with the ordinary laser interferometer. Note that P-V values of the pupil mirrors were measured within an effective area of 4.5~\si{mm} in diameter.

All mirrors except the PO2 meet the requirement for P-V shape error of less than 300~\si{nm} as summarised in Table~\ref{tab:shape_error}. The large shape error of the PO2 comes from its thickness of $\sim$3~\si{mm} due to the limited total size of the IFU; we fixed the workpiece to a jig during the process and slight deformation of the shape may have occurred when it was removed from the jig after finishing. We performed a ray-tracing analysis to evaluate the degradation of image quality caused by the shape error of the PO2, where the measured shape error of the PO2 was simulated as vertical astigmatism, and confirmed that only a few \% degradation of the RMS spot radius at the final image plane of the IFU is caused, which is acceptable. 

\begin{table}[tb]
    \centering
    \caption{Measured P-V shape error of mirrors.}
    \label{tab:shape_error}
    \begin{tabular}{lrrrrr} \toprule
        \rule[-1ex]{0pt}{3.5ex} & \multicolumn{1}{c}{PO0} & \multicolumn{1}{c}{PO2} & \multicolumn{1}{c}{Slice mirrors} & \multicolumn{1}{c}{Pupil mirrors} & \multicolumn{1}{c}{Slit mirrors} \\ \midrule
        \rule[-1ex]{0pt}{3.5ex}Mean [nm] & 240 & 507 & 138 & 211 & 169 \\
        \rule[-1ex]{0pt}{3.5ex}Min [nm] & - & - & 86 & 155 & 117 \\
        \rule[-1ex]{0pt}{3.5ex}Max [nm] & - & - & 228 & 297 & 277 \\ \bottomrule
    \end{tabular}
\end{table}

\subsection{Assembly}
\label{subsec:assemlby}
The optical elements of the IFU should be assembled within tolerances listed in Table~\ref{tab:assembly_tolerance}. They were derived from a tolerance analysis using Zemax, where sensitivity of each degree of freedom to throughput, image quality at the slice-mirror array and at the detector, and image displacement at the detector was evaluated first. Then, a Monte Carlo simulation with 1000 trials was performed to evaluate the effect of combined errors. Within the tolerances in Table~\ref{tab:assembly_tolerance}, 99.7\% of the trials show the throughput degradation of less than 10\%, the variations of the RMS spot radius at the slice-mirror array and at the detector of less than 20\% and 10\% respectively, and the image misalignment at the detector of a few pixels.

\begin{table}[tb]
    \centering
    \caption{Installation tolerances for the optical elements of the IFU. Coordinates are illustrated in Figure~\ref{fig:swimsifu_optics}, except for the PO1, which has tilted coordinates with the Y axis aligned with the optical axis at the PO1.}
    \label{tab:assembly_tolerance}
    \begin{tabular}{lrrrrrr} \toprule
        \rule[-1ex]{0pt}{3.5ex} & \multicolumn{1}{c}{Shift X} & \multicolumn{1}{c}{Shift Y} & \multicolumn{1}{c}{Shift Z} & \multicolumn{1}{c}{Tilt X} & \multicolumn{1}{c}{Tilt Y} & \multicolumn{1}{c}{Tilt Z} \\
        \rule[-1ex]{0pt}{1.5ex} & \multicolumn{1}{c}{[\si{\um}]} & \multicolumn{1}{c}{[\si{\um}]} & \multicolumn{1}{c}{[\si{\um}]} & \multicolumn{1}{c}{[\si{arcmin}]} & \multicolumn{1}{c}{[\si{arcmin}]} & \multicolumn{1}{c}{[\si{arcmin}]} \\ \midrule
        \rule[-1ex]{0pt}{3.5ex}PO0$+$slice-mirror array & 10 & 20 & 10 & 0.9 & 0.8 & 0.7 \\
        \rule[-1ex]{0pt}{3.5ex}PO1 & 10 & 20 & 20 & 2.2 & 1.6 & 1.6 \\
        \rule[-1ex]{0pt}{3.5ex}PO2 & 20 & 20 & 10 & 0.8 & 0.4 & 0.4 \\
        \rule[-1ex]{0pt}{3.5ex}Pupil-mirror array & 20 & 20 & 5 & 0.4 & 0.4 & 0.4 \\
        \rule[-1ex]{0pt}{3.5ex}Slit-mirror array & 10 & 20 & 10 & 1.1 & 0.4 & 0.4 \\ \bottomrule
    \end{tabular}
\end{table}

\begin{figure}[tb]
  \centering
  \includegraphics[keepaspectratio,  width=0.9\hsize]{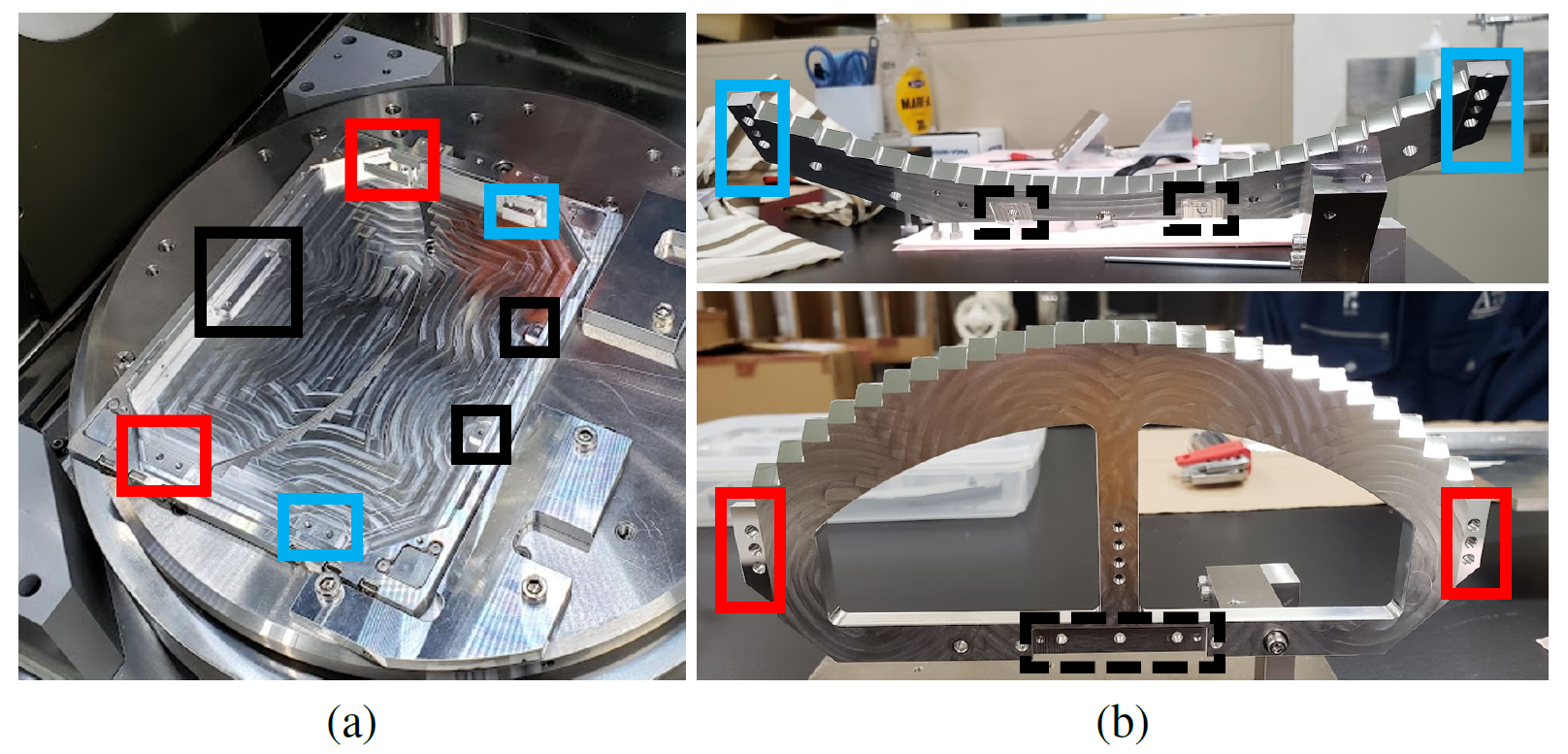}
  \caption{Ultra-precision cutting for assembly. Solid boxes show machined surfaces. (a) Mounting surfaces of the base plate. Blue boxes show mounting surfaces for the pupil-mirror array and red boxes for the slit-mirror array. Black boxes show machined surfaces for unmachined surfaces of the pupil-mirror array and the slit-mirror array shown in dashed black boxes in (b). (b) Fixing flanges of the pupil-mirror array (top) and the slit-mirror array (bottom).}
  \label{fig:ultraprecision_cutting_assembly}
\end{figure}

The ultra-precision cutting was also used in the assembly process to realize simple and precise assembly. First, mounting surfaces of the base plate were machined by ultra-precision cutting to control their height with a few \si{\um} accuracy (Figure~\ref{fig:ultraprecision_cutting_assembly} (a)). Fixing flanges of the pupil-mirror array and the slit-mirror array were also machined in the same operation as the mirrors (Figure~\ref{fig:ultraprecision_cutting_assembly} (b)), so relative positions and orientations between the mirrors and the base plate are guaranteed. For unmachined fixing surfaces of the pupil-mirror array and the slit-mirror array and for the optical elements whose fixing flanges cannot be machined, shims whose thickness was controlled by ultra-precision cutting were used for accurate positioning. Positioning pins were used to fix the optical elements in the plane of the machined surfaces or shims with an accuracy of 10--20~\si{\um}. A lens holder of the PO1 was fixed to fixing flanges next to the PO0 mirror.

\begin{figure}[tb]
    \centering
    \includegraphics[keepaspectratio,  width=0.8\hsize]{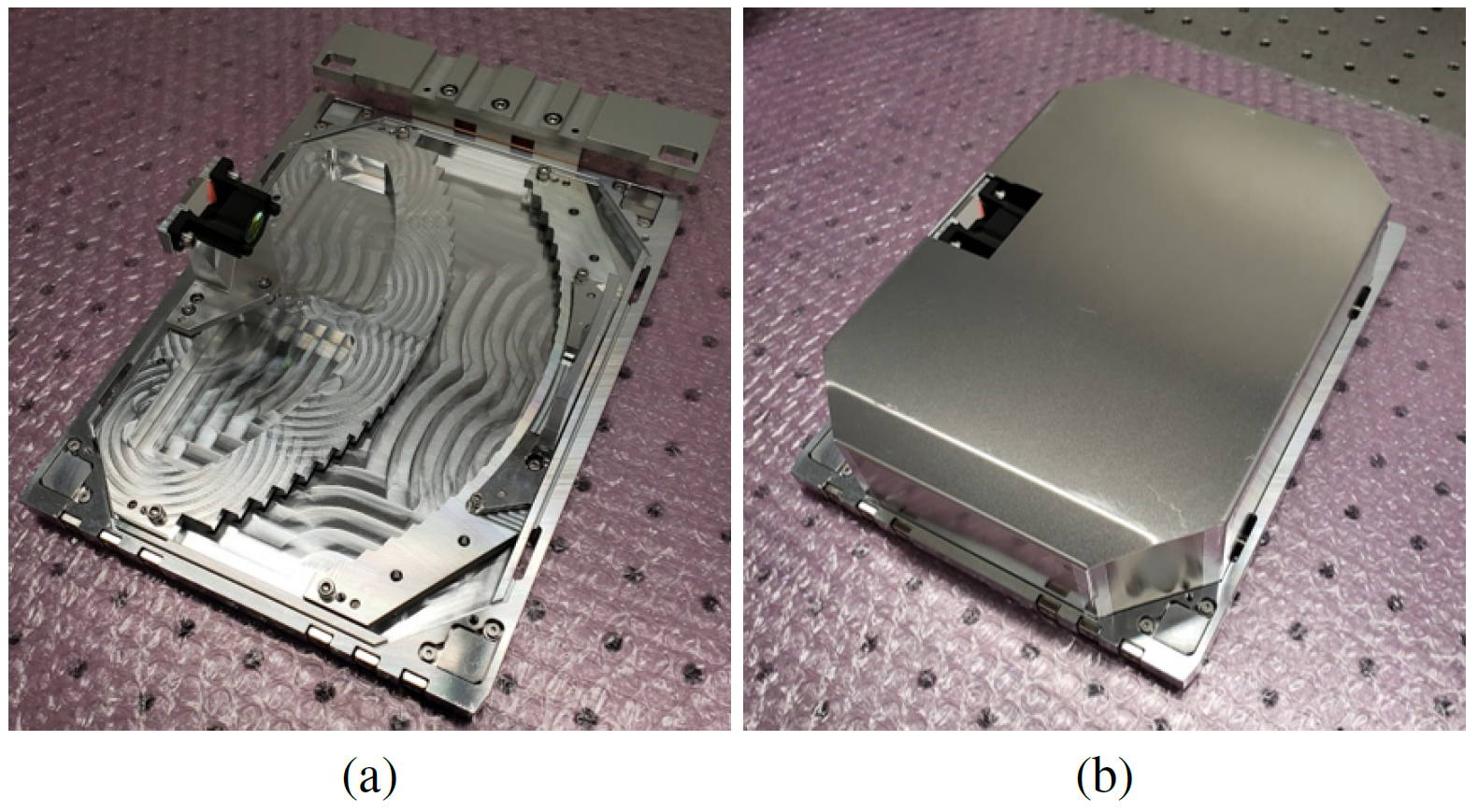}
    \caption{Completed IFU. (a) Without a cover. (b) With a cover.}
    \label{fig:assembly}
\end{figure}

As a result, the assembly of the IFU was completed with only one adjustment of the PO1 lens holder. The completed IFU is shown in Figure~\ref{fig:assembly}. The assembly of the SWIMS-IFU using ultra-precision cutting is simple and does not require precise adjustment. The concept is applicable to not only IFUs but also to a variety of optical units with complex designs.

\section{Laboratory Performance Tests}
\label{sec: lab_evaluation}
\subsection{Vignetting}
\label{subsec: vignetting}
There are two vignetting points for the SWIMS-IFU: One is vignetting of the pupil images at the edges of the FoV by an aperture of the PO1, and the other is vignetting of the FoV by a lens holder of the PO1 (also refer to Figure \ref{fig:swimsifu_mechanics}).

\begin{figure}[tb]
  \centering
  \includegraphics[keepaspectratio,  width=0.9\hsize]{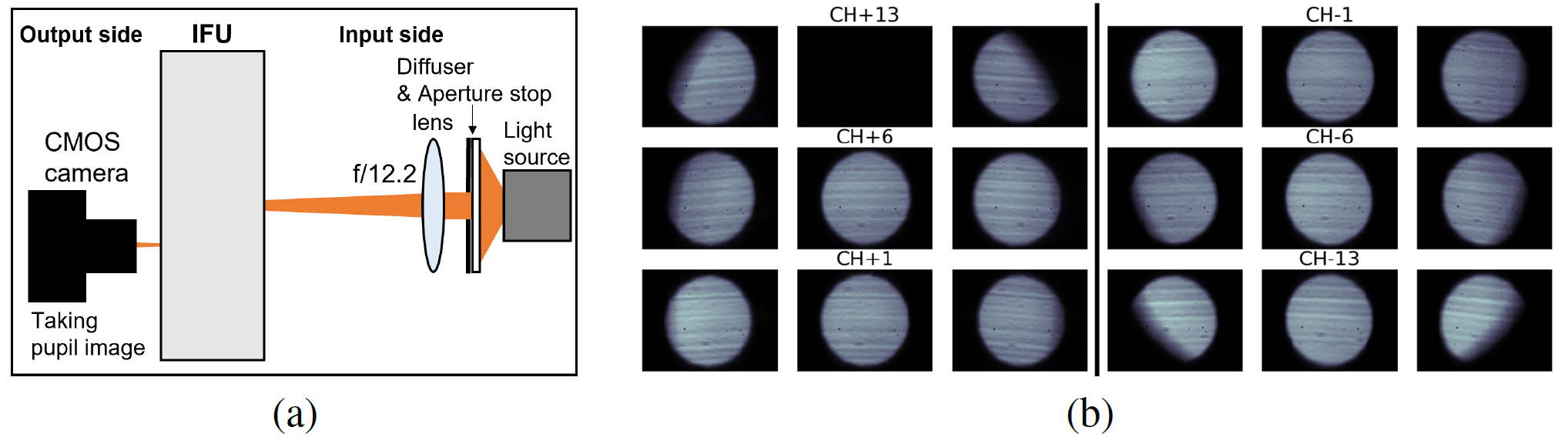}
  \caption{(a) Configuration to measure vignetting of the pupil images by the aperture of the PO1. (b) Observed pupil images at the center and both edges of CH$+$13, $+$6, $+$1, $-$1, $-$6, and $-$13. The pupil image of the central part of CH$+$13 is missing due to the vignetting of the FoV.}
  \label{fig:vignetting_pupil}
\end{figure}

To measure the vignetting of the pupil images by the aperture of the PO1, we set a light source in the visible wavelength with an input F-ratio of 12.2 at the input side of the IFU to simulate incident light of the Subaru telescope and took an illumination pattern at the pupil-mirror array from the output side by a CMOS camera as shown in Figure~\ref{fig:vignetting_pupil} (a). Figure~\ref{fig:vignetting_pupil} (b) shows the observed illumination patterns for various channels with different positions in it. The vignetting of the pupil images at the edges of the FoV is larger than at the center, resulting in lower throughput there. This is caused by an insufficient effective clear aperture diameter of the PO1 lens due to space limitations.

\begin{figure}[tb]
  \centering
  \includegraphics[keepaspectratio,  width=0.7\hsize]{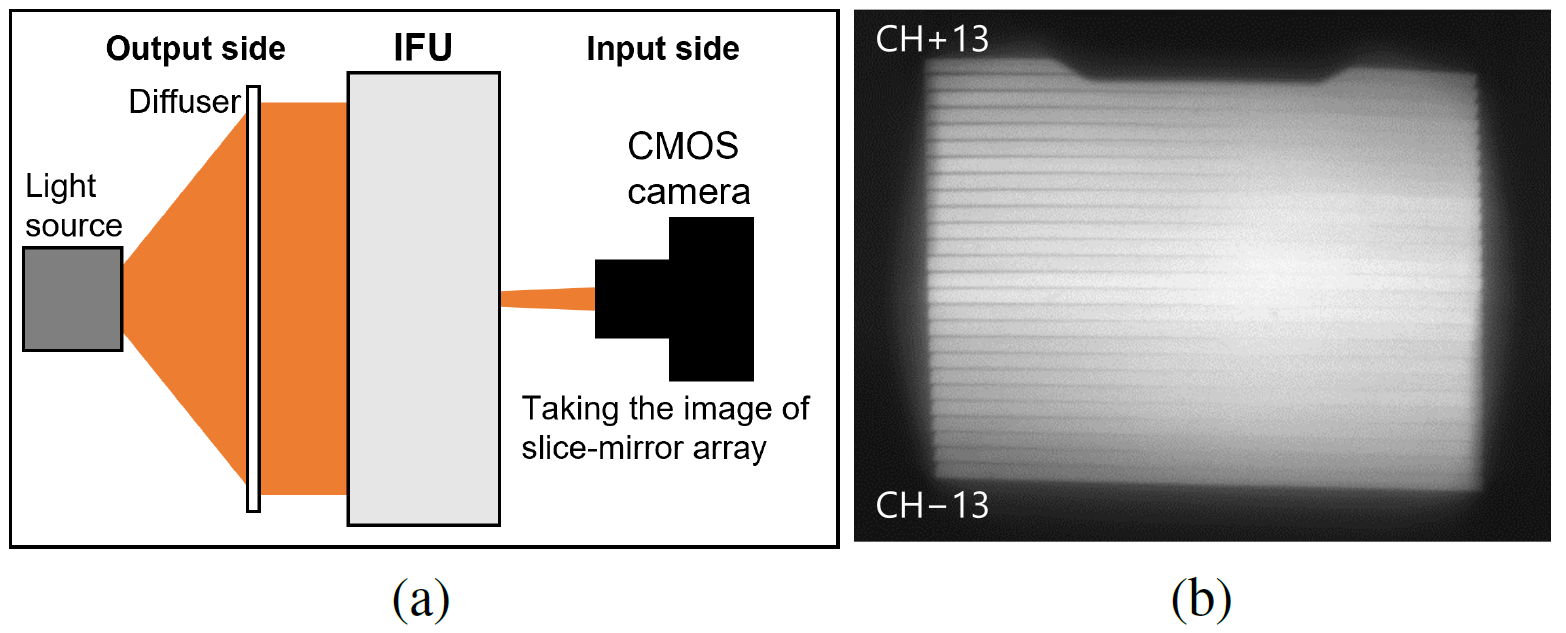}
  \caption{(a) Configuration to measure vignetting of the FoV by the lens holder of the PO1. (b) Observed image of the slice-mirror array. There is vignetting at the top of the image, affecting the central part of CH$+$13 and CH$+$12.}
  \label{fig:vignetting_fov}
\end{figure}

To measure the vignetting of the FoV by the lens holder of the PO1, we set a diffused visible light source at the output side of the IFU shown in Figure~\ref{fig:vignetting_fov} (a) and took an illumination pattern at the slice-mirror array from the input side by a CMOS camera. Figure~\ref{fig:vignetting_fov} (b) shows the observed illumination pattern at the slice-mirror array. It can be seen that the centers of CH$+$13 and CH$+$12 are vignetted, which is caused by obstruction of the light between the slice-mirror array and the PO2 by a lens holder of the PO1 lens.

\subsection{Image Quality}
\label{sec:lab_image_quality}
To measure image quality, we again set the light source with F/12.2 and took the pseudo-slit images of all 26 channels by the CMOS camera from the output side of the IFU. The setup for this test was the same as Figure~\ref{fig:vignetting_pupil} (a), except that the CMOS camera focused on the pseudo-slit image instead of the pupil image.

To evaluate the image quality, cross sections of each pseudo-slit image were fitted with a convolution function of a top-hat function with the designed width and a Gaussian function. The analysis was done for the center and both edges of each pseudo-slit image.

\begin{figure}[tb]
    \centering
    \includegraphics[keepaspectratio, width=\hsize]{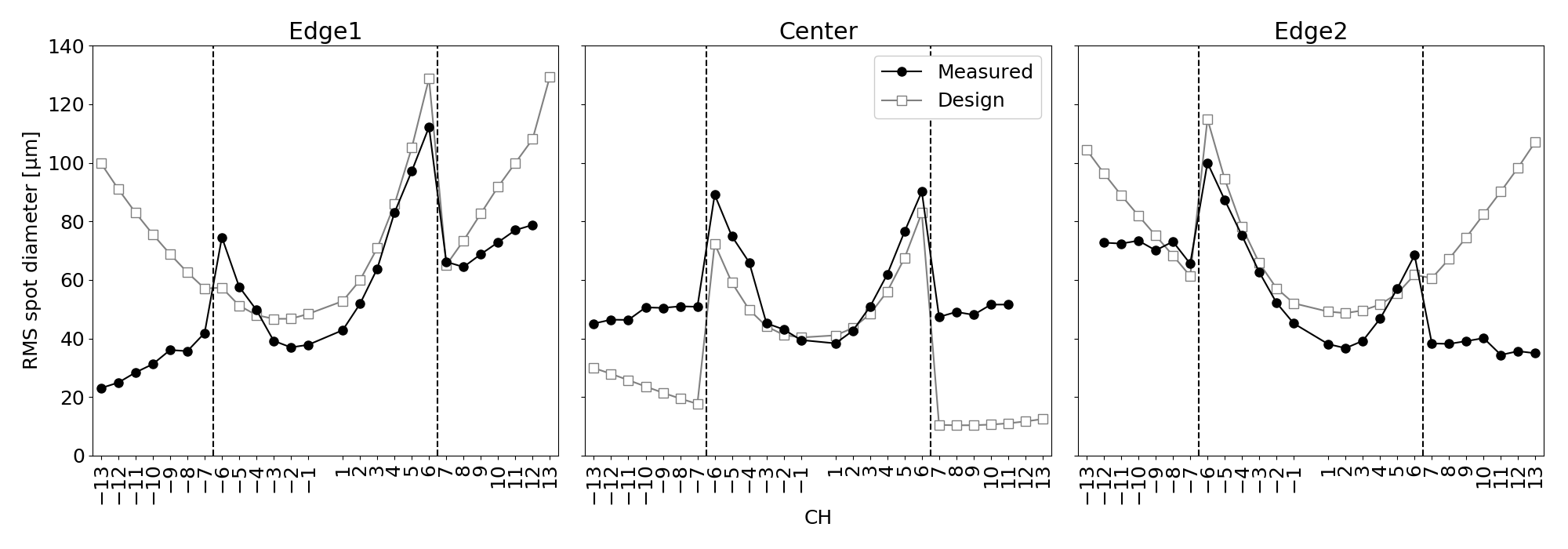}
    \caption{Measured image quality of the pupil mirrors at the center and both edges of the pseudo-slit image and ray-tracing simulated values in 700~\si{nm}. Black circles show the measured value and white squares show the ray-tracing simulated values. Dashed vertical lines show the boundaries between channels with a spherical pupil mirror and with an off-axis ellipsoidal pupil mirror. Measured values for the center of CH$+$12 and CH$+$13 are missing due to vignetting of the FoV, and measured values for edge1 of CH$+$13 and edge2 of CH$-$13 are missing due to the configuration of the measurement.}
    \label{fig:measured_image_quality}
\end{figure}

Figure~\ref{fig:measured_image_quality} shows double the standard deviations of the Gaussian functions, which we interpret as RMS spot diameters, for all channels, as well as the RMS spot diameters from a ray-tracing simulation at the position of the pseudo-slit image. We should note that these values include degradation caused by various elements. The ray-tracing simulation simulates a spot for a point source at infinity degraded by the telescope and all the optical elements of the IFU. On the other hand, the laboratory measurements consider only the degradation by the pupil mirrors. Thus, some measurements at the edge of the outer channels show an RMS spot diameter smaller than the ray-tracing simulation. Still, the overall trend of the measurement is consistent with the ray-tracing simulation, demonstrating the effect of the off-axis ellipsoidal pupil mirrors for the outer channels. 

\section{Engineering Observation}
\label{sec: eng_observation}
Engineering observation at Subaru telescope was carried out on the first half of the night of March 27, 2022, and on the first half of the night of December 2, 2022. Through these engineering observations, we evaluated the on-sky performance of the SWIMS-IFU and demonstrated that the development utilizing ultra-precision cutting can complete the IFU as designed. This allows us to improve performance by revising the IFU optics and mechanics, and to quickly start the operation of the SWIMS-IFU at the TAO 6.5~{\si{m}} telescope.

\begin{figure}[tb]
    \centering
   \includegraphics[keepaspectratio,  width=\hsize]{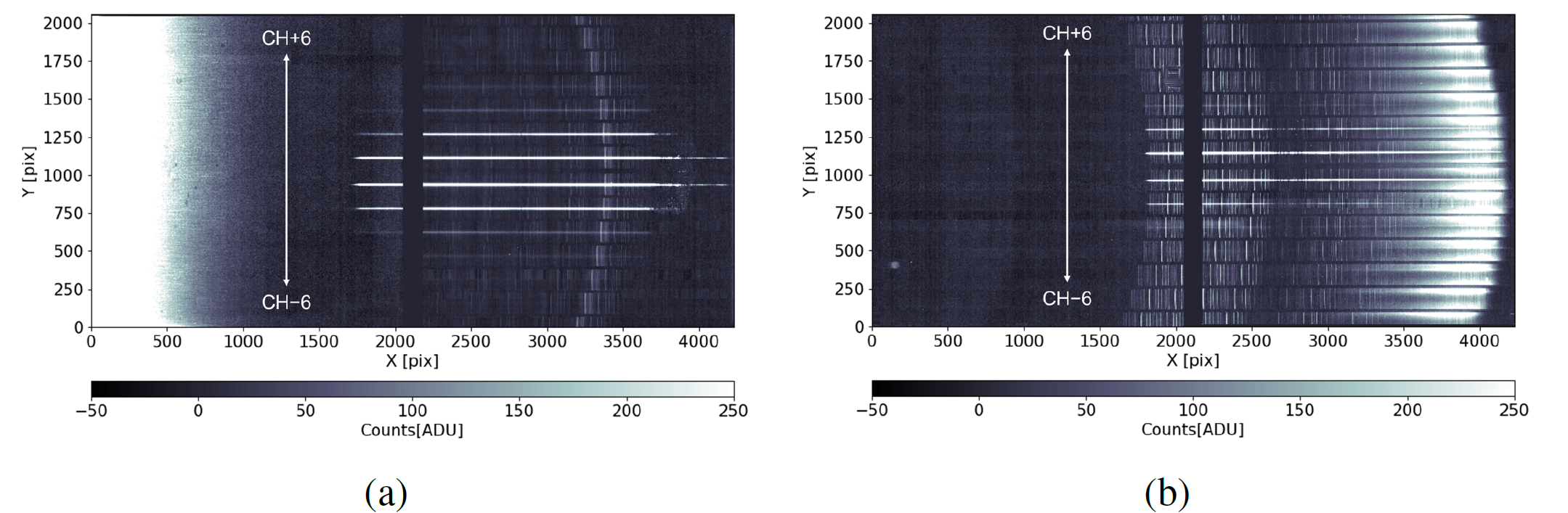}
    \caption{Observed two-dimensional spectra of a standard star, HIP51684, of (a) the blue arm covering from 0.9~\si{\um} to 1.45~\si{\um}, and (b) the red arm from 1.45~\si{\um} to 2.5~\si{\um}.}
    \label{fig:observed_spectra}
\end{figure}

\begin{figure}[tb]
    \centering
    \includegraphics[keepaspectratio,  width=0.9\hsize]{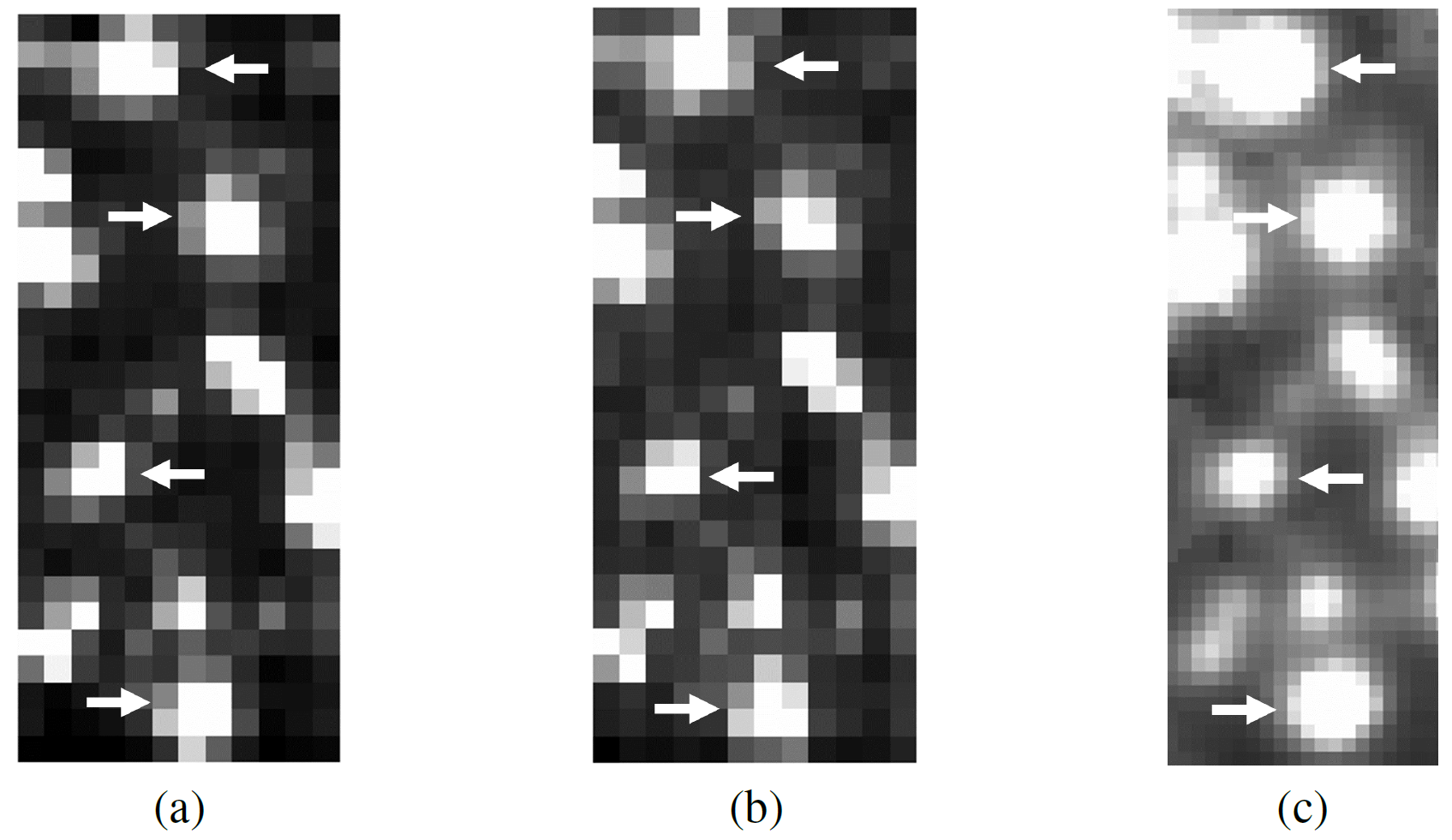}
    \caption{Comparison of images of M3. White arrows indicate stars used in the scale measurement. Reconstructed images of the IFU are binned every 5 pixels in the vertical direction. (a) Reconstructed $J$-band image of the FoV of the IFU. (b) Reconstructed $K_S$-band image of the FoV of the IFU. (c) Pan-STARRS $y$-band image.}
    \label{fig:m3_image}
\end{figure}

Figure~\ref{fig:observed_spectra} shows observed two-dimensional spectra of a standard star, HIP51684, placed at the center of the FoV with integration times of 20~\si{s} for the blue arm and 10~\si{s} for the red arm of SWIMS. The spectra of the standard star were obtained at nine positions in the FoV: the center and both edges of CH$-$1, CH$-$4, and CH$+$4. Figure \ref{fig:m3_image} (a) and (b) are images of a globular cluster M3 reconstructed from the pseudo-slit images of the IFU with an integration time of 3~\si{s} for $J$- and $K_S$-band filters, respectively. Four pseudo-slit images were obtained in the engineering observation.

\subsection{Spatial Scales and FoV}
The pseudo-slit imaging data of the globular cluster M3 were used for spatial scale measurement. The pseudo-slit images were reconstructed into an image of the FoV of the IFU. Positions of stars in the reconstructed image were compared with those in a $y$-band image of Pan-STARRS\cite{Chambers2016} to measure the slice width and the spatial scale along the slices (Figure~\ref{fig:m3_image}). The positions of the stars were measured with the IRAF imexam task, and the slice width and the spatial scales were computed by the IRAF ccmap task. 

Table~\ref{tab:spatial_scale} shows the mean and standard deviation of measurements for each filter, as well as designed values. Both the slice width and the spatial scale for each filter are almost consistent with the designed values. With the measured slice width and the spatial scale, the FoV of the SWIMS-IFU becomes 13.4 $\times$ 4.8~\si{arcsec^2}.

\begin{table}[tb]
    \centering
    \caption{Measured and designed spatial scales of the IFU.}
    \label{tab:spatial_scale}
    \begin{tabular}{lcc} \toprule
        \rule[-1ex]{0pt}{3.5ex} & \multicolumn{1}{c}{Slice width [\si{arcsec}]} & \multicolumn{1}{c}{Spatial scale [\si{arcsec.pix^{-1}}]} \\ \midrule
        \rule[-1ex]{0pt}{3.5ex}Blue arm ($J$) & 0.403$\pm$0.009 & 0.0950$\pm$0.0003 \\
        \rule[-1ex]{0pt}{3.5ex}Red arm ($K_S$) & 0.398$\pm$0.023 & 0.0962$\pm$0.0002 \\ 
        \rule[-1ex]{0pt}{3.5ex}Design & \multicolumn{1}{l}{\hspace{16.5pt}0.390} & \multicolumn{1}{l}{\hspace{41.5pt}0.0960} \\ \bottomrule
    \end{tabular}
\end{table}

\subsection{Wavelength Scales and Coverage}
Wavelength calibration was carried out using the night sky OH emission lines\cite{Rousselot2000} in the images of the standard star. Pixel scales in dispersion direction were measured to be 2.47~\si{\AA.pix^{-1}} for the blue arm and 4.67~\si{\AA.pix^{-1}} for the red arm. With the measured pixel scales, observed spectra cover wavelength ranges of 0.85--1.48~\si{\um} for the blue arm and 1.42--2.56~\si{\um} for the red arm. It should be noted that a gap between the two detectors, as seen in Figure~\ref{fig:observed_spectra}, corresponds to $\sim$ 320~\si{\AA} width in the range of 9420--10070~\si{\AA} for the blue arm and $\sim$ 585~\si{\AA} width in 15510--16770~\si{\AA} for the red arm, affecting the observed spectra.

\subsection{Image Quality}
We evaluated image quality using the size of the image of the standard star HIP51684 at the slice-mirror array and at the SWIMS detector. Due to time constraints of the engineering observations, the image quality measurements were made for some of the 12 available channels. However, since the laboratory performance test in Section~{\ref{sec:lab_image_quality}} shows the expected results for all channels, it is expected that the other channels not evaluated in the engineering observations will show similar results.

The image quality at the slice-mirror array was evaluated by reconstructing a stellar profile at the slice-mirror array from signal counts of the spectra of the standard star as follows.  For the obtained spectra, several wavelengths that were not affected by atmospheric absorption or night sky emissions were selected, and signal counts of the standard star at the wavelengths were extracted from the 5--6 channels. The extracted signal counts were arranged in the order of the slice mirrors to yield a stellar profile in the direction perpendicular to the slice mirrors. The stellar profile was fitted with a Gaussian function and an FWHM of the Gaussian function was used to evaluate the size of the standard star image using the slice width of 0.4~\si{arcsec}. The evaluations were performed using data where the standard star was introduced at the nine positions of the FoV of the IFU. 

\begin{figure}[tb]
    \centering
    \includegraphics[keepaspectratio, width=0.6\hsize]{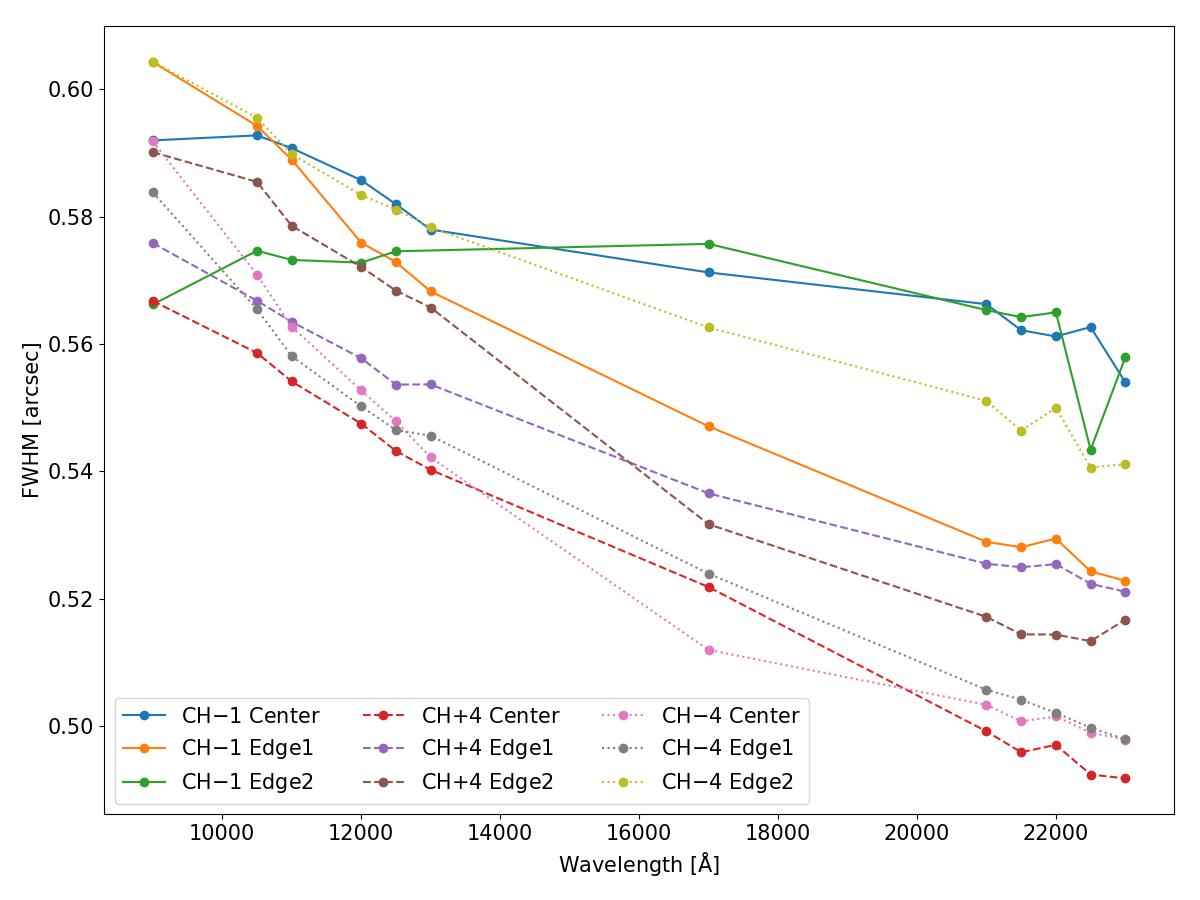}
    \caption{FWHM of the standard star at the slice mirror as a function of wavelength at the nine positions in the IFU FoV.}
    \label{fig:fwhm_at_slice}
\end{figure}

Figure~\ref{fig:fwhm_at_slice} shows the FWHMs at the slice-mirror array as a function of wavelength. The resulting FWHM was 0.49--0.60~\si{arcsec}, which is consistent with an expected value of 0.50--0.53~\si{arcsec} calculated as a root sum square of the image quality from a ray-tracing simulation of 0.205~\si{arcsec} and measured seeing sizes of 0.49~\si{arcsec} in $Y$-band and 0.45~\si{arcsec} in $H$-band.

The image quality at the detectors was evaluated by fitting spatial extent of the standard star spectra observed in CH$\pm$1 and CH$\pm$4 along the slice length direction with a Gaussian function at wavelengths of 1.0~\si{\um} (the central wavelength of the $Y$-band), 1.2~\si{\um} ($J$-band), 1.6~\si{\um} ($H$-band), and 2.1~\si{\um} ($K_S$-band). The Gaussian fitting was done by the IRAF imexam task and the pixel scales in Table~\ref{tab:spatial_scale} were used. Table~\ref{tab:fwhm_at_detector} shows measured FWHMs. They range 0.46--0.65~\si{arcsec} depending on the wavelength, which agrees with an expected value of 0.53--0.56~\si{arcsec} calculated as a root sum square of the image quality from a ray-tracing simulation of 0.268~\si{arcsec} and the measured seeing sizes of 0.49~\si{arcsec} in $Y$-band and 0.45~\si{arcsec} in $H$-band. There is no significant difference between the channels. It should be noted that a gap in the FWHM between the blue ($Y$- and $J$-band) and the red ($H$- and $K_S$-band) arms is caused by SWIMS optics and is not due to the IFU.

\begin{table}[tb]
    \centering
    \caption{FWHM of the standard star in the spatial direction at the detectors.}
    \label{tab:fwhm_at_detector}
    \begin{tabular}{lcccc} \toprule
        \rule[-1ex]{0pt}{3.5ex}CH & FWHM $Y$ [\si{arcsec}] & FWHM $J$ [\si{arcsec}] & FWHM $H$ [\si{arcsec}] & FWHM $K_S$ [\si{arcsec}] \\ \bottomrule
        \rule[-1ex]{0pt}{3.5ex}$-$1 & 0.621 & 0.610 & 0.473 & 0.473 \\
        \rule[-1ex]{0pt}{3.5ex}$+$1 & 0.616 & 0.605 & 0.498 & 0.481 \\
        \rule[-1ex]{0pt}{3.5ex}$-$4 & 0.645 & 0.630 & 0.460 & 0.457 \\
        \rule[-1ex]{0pt}{3.5ex}$+$4 & 0.615 & 0.598 & 0.483 & 0.479 \\ \bottomrule
    \end{tabular}
\end{table}

\subsection{Throughput}
Throughput of the SWIMS-IFU was evaluated as a ratio of counts of dome flat data with the IFU to that without the IFU. The dome flat data were obtained in $Y$-, $J$-, $H$-, and $K_S$-band for imaging mode and in spectroscopy mode. For the spectroscopy mode, dome flat data without the IFU were obtained using a slit mask with slits in the same positions and widths as those of the IFU pseudo-slit images.

\begin{figure}[tb]
    \centering
    \includegraphics[keepaspectratio, width=0.8\hsize]{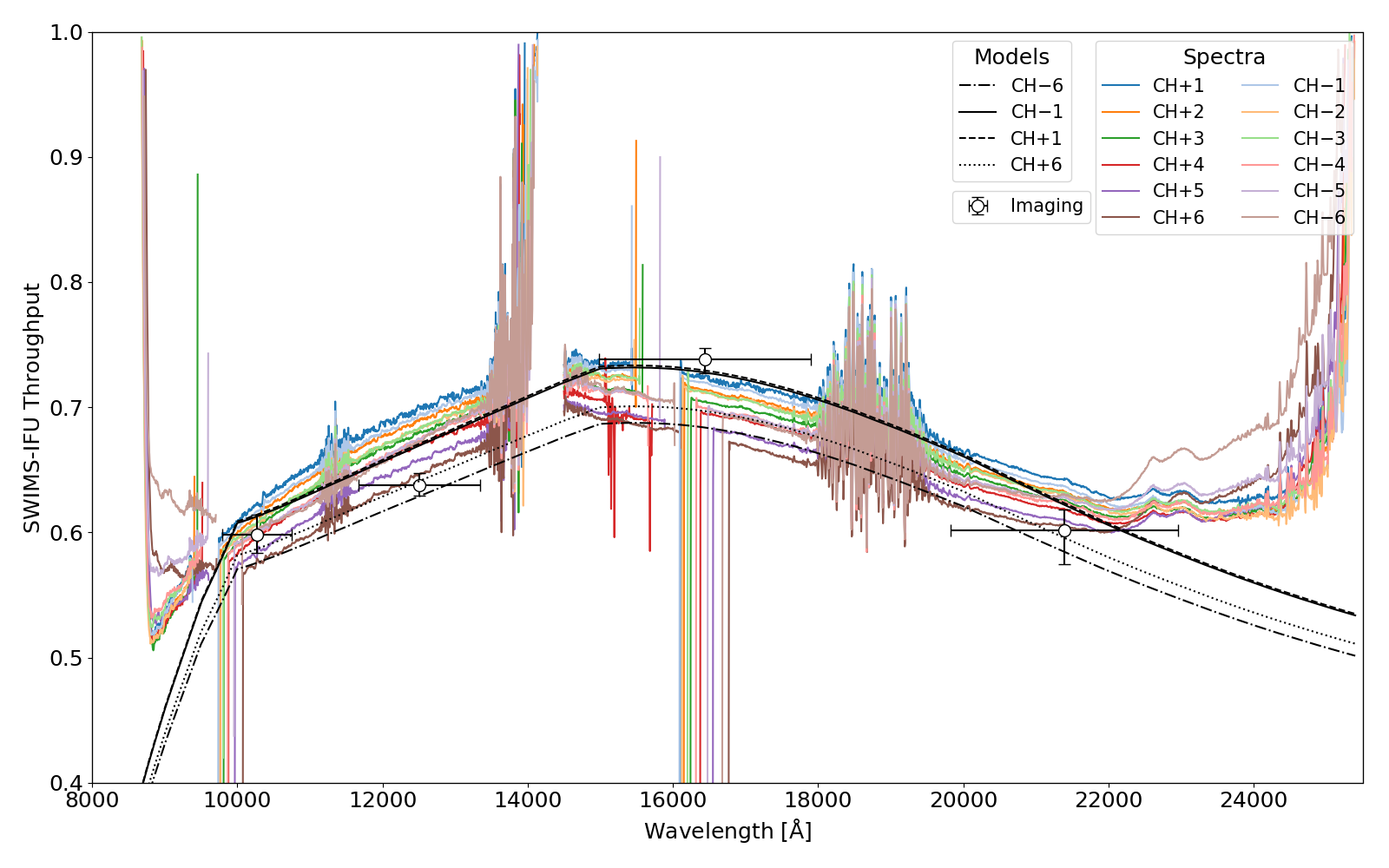}
    \caption{Throughput of the SWIMS-IFU. Open circles show the average throughput of twelve channels by $Y$-, $J$-, $H$-, and $K_S$-band imaging. Error bars show the wavelength range of each filter and the range from minimum to maximum values for the twelve channels. Colored lines show the throughput from the spectroscopy data. Black lines show the throughput models calculated from the scattering loss by the measured mirror surface roughness, aluminum mirror reflectivity, internal transmittance and reflectivity of anti-reflection coating of the PO1 lens, and losses by vignetting for CH$\pm$1 and CH$\pm$6.}
    \label{fig:throughput}
\end{figure}

Figure~\ref{fig:throughput} shows the results, which range from 50\% to 75\% depending on the wavelength and channel. The results are almost consistent between the imaging and the spectroscopy modes. In Figure~\ref{fig:throughput}, we also show throughput models including scattering losses by the measured surface roughness of manufactured mirrors using Equation~(\ref{eq:roughness}), reflectivity of uncoated aluminum mirrors\cite{NAOJ2022}, internal transmittance and reflectivity of anti-reflection coating of the PO1, and losses by vignetting for CH$\pm$1 and CH$\pm$6. The throughput models for the other channels should be between the models of CH$\pm$1 and CH$\pm$6 because the loss by vignetting is between those of CH$\pm$1 and CH$\pm$6.  We can see that the measured values are consistent with the throughput models. The throughput is dominated by the reflectivity of uncoated aluminum mirrors in shorter wavelengths and the reflectivity of the anti-reflection coating of the PO1 in longer wavelengths. The vignetting described in Section \ref{subsec: vignetting} also affects the throughput over the entire wavelength range.

\subsection{Stray Light}
\begin{figure}
    \centering
    \includegraphics[keepaspectratio, width=0.7\hsize]{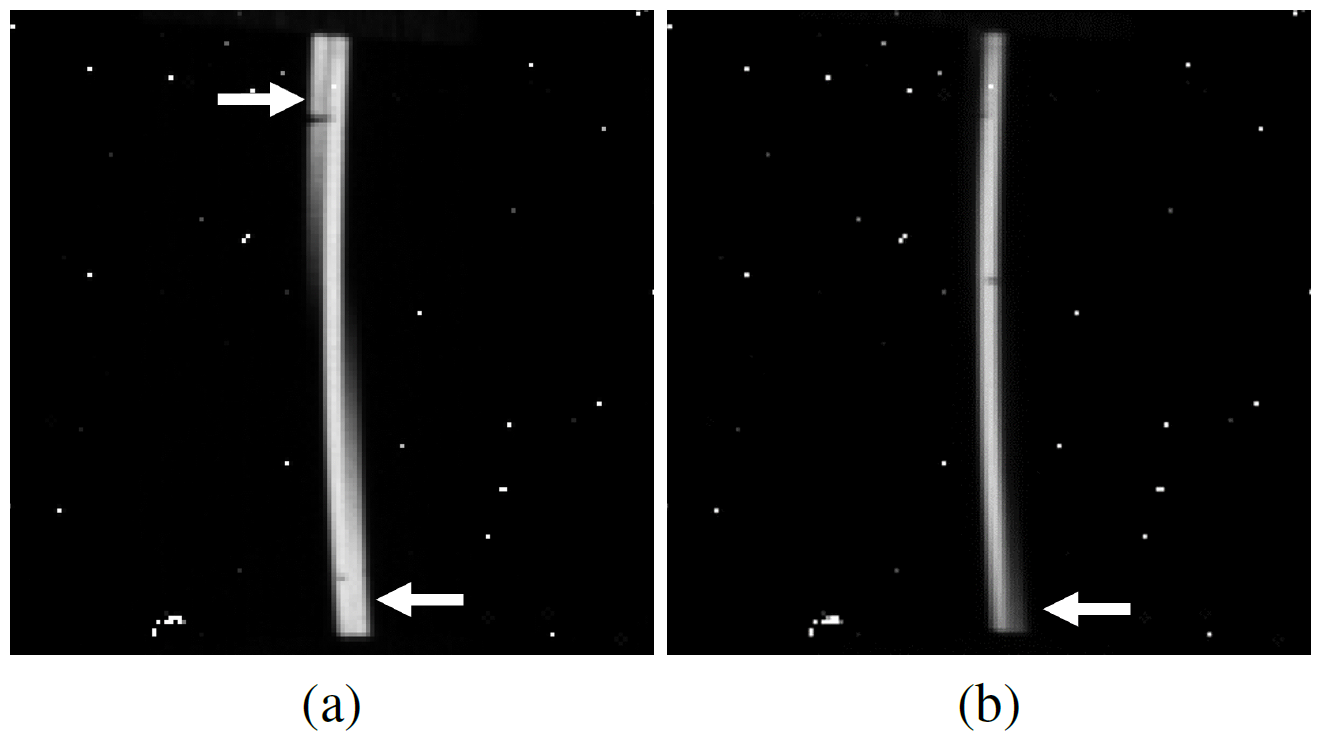}
    \caption{Stray light observed in the pseudo-slit image of CH$-$1 in the $K_S$-band, (a) before modification of the entrance aperture of the SWIMS-IFU and (b) after modification. White arrows indicate the position of the stray light.}
    \label{fig:stray_light}
\end{figure}

During the first engineering observation, we observed stray light affecting wavelengths longer than 2.1~\si{\um} (Figure~\ref{fig:stray_light} (a)). Ray-tracing analysis revealed that the stray light was caused by thermal emission from a baffle around the Cassegrain hole of the Subaru telescope, which enters the IFU at different incident angles and paths from the designed light and is reflected by a slice mirror and then enters a pupil mirror in an adjacent channel. Based on the results of the ray-tracing analysis, we narrowed the entrance aperture of the IFU in the slice length direction from 14.5~\si{mm} to 8.5~\si{mm} to suppress the stray light as shown in Figure\ref{fig:stray_light} (b).

Even though we could reduce the thermal stray light, such stray light persists in the current optical design of the SWIMS-IFU and modification in the optical design is necessary to eliminate the stray light in the future upgrade for the TAO 6.5~\si{m} telescope. One solution is to narrow the FoV and thus the shape of the entrance aperture of the IFU to prevent the stray light from entering the IFU. Another is to change the arrangement of the pupil mirrors, especially in the vertical direction, to prevent the stray light from entering the adjacent channel, like SPIFFI of VLT/SINFONI\cite{Eisenhauer2000}. 

\section{Conclusions}
\label{sec: conclusions}
We have developed an image-slicer type integral field unit with a larger FoV for a near-infrared spectrograph, SWIMS. The FoV of 13.5 $\times$ 10.4~\si{arcsec^2} that is larger than existing near-infrared IFS instruments was realized by matching the width of the slice mirrors to a seeing size at the Subaru telescope of 0.4~\si{arcsec}. We also realized a compact optical design to fit within a volume of 235 $\times$ 170 $\times$ 55~\si{mm^3} by using off-axis ellipsoidal pupil mirrors for the outer channels and reducing aberrations.

Its complicated optical elements were fabricated by ultra-precision cutting; two types of flat mirrors in the PO0$+$slice-mirror array were simultaneously fabricated using dual-mounted tools and free-form surfaces in the pupil-mirror array and the slit-mirror array were fabricated by ball end milling with a small-radius tool. The finished mirrors meet our requirements of RMS surface roughness of less than 10~\si{nm} and P-V shape error of less than 300~\si{nm}. We also used the ultra-precision cutting to machine mounting surfaces and custom-made shims for easy assembly. As a result, the assembly was completed with only one adjustment of the PO1 lens. The concept of the easy assembly can be applied to the development of other optical systems. Remaining technical issues are how to measure the shape of aspherical mirrors and how to assess their relative positions.

The main results of the performance evaluation of the completed IFU are:
\begin{itemize}
    \item We confirmed the improvement of image quality by the off-axis ellipsoidal pupil mirrors through the laboratory test. The engineering observation at Subaru telescope showed the FWHMs of 0.49--0.60~\si{arcsec} at the slice-mirror array and 0.46--0.65~\si{arcsec} at the detectors, which are consistent with their expected values.
    \item The throughput of the IFU was 50--75\% as designed. It is dominated by the reflectivity of uncoated aluminum mirrors in shorter wavelengths and the reflectivity of the anti-reflection coating of the PO1 in longer wavelengths.
    \item Thermal emission from the baffle of the telescope was observed as stray light. It was confirmed to be suppressed by narrowing the entrance aperture of the IFU, however, modification of arrangement of the pupil mirrors is necessary to remove it completely.
\end{itemize}

The SWIMS-IFU will be upgraded for the observation at the TAO 6.5~\si{m} telescope. To improve the vignetting, we will remove the PO1 lens from the optical design and realize the required magnification by adopting a pick-off mirror with power. Additional aberration caused by the new pick-off mirror will be corrected using aspherical mirrors. This update will also improve the low throughput in the longer wavelengths. The low throughput in the shorter wavelengths can be improved by coating the mirrors with gold. To remove the thermal stray light, we have to create a more complicated slice-mirror array to modify the arrangement of the pupil images. The fabrication of the new slice-mirror array will be a new technical challenge.

\section*{Code, Data, and Materials Availability}
The data that support the findings of this article are not publicly available. They can be requested by contacting the corresponding author.

\section*{Acknowledgments}
We appreciate the support of Subaru telescope for the engineering observation. The development activities are supported by the Advanced Technology Center, NAOJ, and the Advanced Manufacturing Support Team, RIKEN. This work made use of iraf\cite{Tody1986, Tody1993} and pyraf\cite{ScienceSoftwareBranchatSTScI2012}. This work made use of Astropy (\url{http://www.astropy.org}) a community-developed core Python package and an ecosystem of tools and resources for astronomy \cite{AstropyCollaboration2013, AstropyCollaboration2018, AstropyCollaboration2022}. The Pan-STARRS1 Surveys (PS1) and the PS1 public science archive have been made possible through contributions by the Institute for Astronomy, the University of Hawaii, the Pan-STARRS Project Office, the Max-Planck Society and its participating institutes, the Max Planck Institute for Astronomy, Heidelberg and the Max Planck Institute for Extraterrestrial Physics, Garching, The Johns Hopkins University, Durham University, the University of Edinburgh, the Queen's University Belfast, the Harvard-Smithsonian Center for Astrophysics, the Las Cumbres Observatory Global Telescope Network Incorporated, the National Central University of Taiwan, the Space Telescope Science Institute, the National Aeronautics and Space Administration under Grant No. NNX08AR22G issued through the Planetary Science Division of the NASA Science Mission Directorate, the National Science Foundation Grant No. AST-1238877, the University of Maryland, Eotvos Lorand University (ELTE), the Los Alamos National Laboratory, and the Gordon and Betty Moore Foundation. This work was supported by JSPS KAKENHI Grant Numbers JP23540261, JP15H02062, JP20H00171, JP14J06780, JP20J21493. This work was supported by the Research Coordination Committee, National Astronomical Observatory of Japan (NAOJ). KK was supported as a Research Fellow of Japan Society for the Promotion of Science. This work is based on ``Development status of a near-infrared integral field unit SWIMS-IFU'', Proceedings of the SPIE, 12188, 121882V (2022)\cite{Kushibiki2022}, ``Diamond machining of two-in-one optical element including slice mirror array for near-infrared integral-field spectrograph'', Proceedings of the SPIE, 12188, 121882X (2022)\cite{Takeda2022}, and Ph.D. Thesis of KK of the University of Tokyo (2023).

\bibliography{article}   
\bibliographystyle{spiejour}   


\vspace{2ex}\noindent\textbf{Kosuke Kushibiki} is a project researcher at the Institute of Astronomy, the University of Tokyo. He received his Ph.D. in astronomy from the University of Tokyo in 2023. His current research interests include nearby starburst galaxies, galaxy mergers, and quenching. He has been involved in the development of an integral field unit. He is also working as a member of the TAO project since 2023.

\vspace{1ex}
\noindent Biographies and photographs of the other authors are not available.

\end{document}